\pdfoutput=1

\documentclass[11pt,twoside,a4paper,cmspaper,final,collab]{cms-tdr}
\def\svnVersion{74863:76252}\def\cmsCernNo{2011-085}\def\cmsCernDate{\today}\def\cmsMessage{Submitted to Physical Review D}

\begin{document}\cmsNoteHeader{TOP-10-003}

\hyphenation{had-ron-i-za-tion}
\hyphenation{cal-or-i-me-ter}
\hyphenation{de-vices}
\RCS$Revision: 76252 $
\RCS$HeadURL: svn+ssh://alverson@svn.cern.ch/reps/tdr2/papers/TOP-10-003/trunk/TOP-10-003.tex $
\RCS$Id: TOP-10-003.tex 76252 2011-08-18 14:56:36Z alverson $
\cmsNoteHeader{TOP-10-003} 
\title{Measurement of the \ttbar Production Cross Section in $\Pp\Pp$ Collisions at 7 TeV in Lepton + Jets Events Using $\cPqb$-quark Jet Identification}

\date{\today}

\abstract{
A new measurement of the inclusive production cross section for $\Pp\Pp \to \ttbar$ is performed at a center-of-mass energy of 7\TeV using data collected by the CMS experiment at the LHC. The analysis uses a data sample corresponding to an integrated luminosity of 36\pbinv, and is based on the final state with one isolated, high transverse momentum muon or electron, missing transverse energy, and hadronic jets. The $\ttbar$ content of the selected events is enhanced by requiring the presence of at least one jet consistent with $\cPqb$-quark hadronization. The measured cross section is $150 \pm 9~\mathrm{(stat.)} \pm 17~\mathrm{(syst.)} \pm 6~\mathrm{(lumi.)}\unit{pb}$ and is in agreement with higher-order QCD calculations. The combination of this measurement with a previous CMS result based on dileptons gives $154  \pm 17~\mathrm{(stat.+syst.)} \pm 6~\mathrm{(lumi.)}$\unit{pb}.
}

\hypersetup{%
pdfauthor={CMS Collaboration},%
pdftitle={Measurement of the ttbar Production Cross Section in pp Collisions at 7 TeV in Lepton + Jets Events Using b-quark Jet Identification},%
pdfsubject={CMS},%
pdfkeywords={CMS, LHC, physics, top quark, cross section}%
}

\maketitle 

\section{Introduction}

The top quark was first observed in proton-antiproton collisions at $\sqrt{s} = 1.8\TeV$ at the Fermilab Tevatron collider \cite{top-discovery-cdf,top-discovery-d0}. Since then its properties have been studied by the Tevatron experiments and found to be in agreement with the expectations of the standard model (SM) \cite{top-prospects}. At the Large Hadron Collider (LHC) \cite{lhc} top quark production can be studied in $\Pp\Pp$ collisions at $\sqrt{s}$ = 7\TeV, allowing extended measurements of the top quark properties. A precise measurement of these properties is important as top-quark production may be a background for new physics.

At the LHC top quarks can be produced singly or in pairs. This paper focuses on the study of the $\ttbar$ final state, for which the production cross section has been calculated in next-to-leading-order (NLO) and approximate next-to-next-to-leading-order (NNLO) quantum chromodynamics (QCD) \cite{Aliev:2010zk,Langenfeld:2009wd,Kidonakis:2010dk}. In the SM the top quark decays almost 100\% of the time via the weak process $\cPqt \to \PW\cPqb$. We focus on the $\ttbar$ decays in which one of the two $\PW$ bosons decays hadronically and the other decays leptonically (the semileptonic channel), giving a final state containing an electron or muon, a neutrino, and four jets, two of which come from the hadronization of $\cPqb$ quarks. Here taus are detected only through their semileptonic decays to electrons and muons. The results are based on the analysis of a data sample corresponding to an integrated luminosity of  36\pbinv\cite{CMS_Lumi}, which was recorded by the Compact Muon Solenoid (CMS) experiment between April and November 2010.

We present the results of $\ttbar$ cross section measurements in the muon and electron + jets channels using the subset of the data in which at least one jet has been $\cPqb$ tagged using a displaced secondary vertex algorithm. A profile likelihood method is used to fit the mass distribution of the identified $\cPqb$-decay vertices as a function of the jet and $\cPqb$-tag multiplicities in the event. The analysis is sensitive to the differences between the signal and background processes and also allows the simultaneous fit of the light $(\cPqu, \cPqd, \cPqs)$ and heavy quark $(\cPqb, \cPqc)$ contributions. The main systematic uncertainties are taken into account when maximizing the profile likelihood. This allows for the correct treatment of their correlations and the evaluation of the combined uncertainty. These results complement the CMS dilepton and kinematics-based lepton+jets analyses which are described elsewhere \cite{dilep,lepjets}. The ATLAS experiment has also measured the $\ttbar$ cross section in the dilepton and lepton+jets channels at 7\TeV \cite{ATLAS:2010}. Several cross-check analyses are also performed, which use different analysis techniques (a fit without the use of the profile likelihood, and two measurements based on simple cuts), and different $\cPqb$-tag algorithms (one based on a track impact parameter, and another that uses soft muons).

The central feature of the CMS detector is a superconducting solenoid, 13 m in length and  6~m in diameter, which provides an axial magnetic field of 3.8 T. The inside of the solenoid is outfitted with various particle detection systems. Charged particle trajectories are measured by the silicon pixel and strip tracker, covering $0 < \phi < 2\pi$ in azimuth and $|\eta| < 2.5$, where the pseudorapidity $\eta$ is defined as $\eta = - \ln[\tan(\theta/2)]$, and $\theta$ is the polar angle of the trajectory with respect to the anti-clockwise beam direction. A crystal calorimeter and a brass/scintillator calorimeter surround the tracking volume and provide high resolution energy and direction measurements of electrons, photons, and hadronic jets. Muons are measured in gas-ionization detectors embedded in the steel return yoke outside the solenoid. The detector is nearly hermetic, allowing for energy balance measurements in the plane transverse to the beam direction. A two-tier trigger system selects the most interesting $\Pp\Pp$ collision events for use in physics analysis. A more detailed description of the CMS detector can be found elsewhere~\cite{cms}.

We describe the data and event selection in Section 2 of this paper, followed by a brief description of the modeling of the signal and background processes in Section 3.
Section 4 describes the method used to extract the cross section from the selected events, as well as the calculation of the statistical and systematic uncertainties on the result. The cross-check analyses are discussed briefly in Section 5. Lastly, in Sections 6 and 7 we present a summary of all of the CMS measurements and compare the results with the predictions from QCD.

\section{Event Selection}
\label{sec:EventSelection}

The trigger used to select the data samples for analysis is based on the presence of at least one charged lepton, with either an electron or a muon with a transverse momentum, $\PT > \PT^\mathrm{min}$. Because of increasing maximum instantaneous luminosity, the minimum transverse momentum was varied between 9 and 15\gev for muons and between 10 and 22\gev for electrons in order to maintain a reasonable trigger rate.
The same data are used both for signal selection and for the study of the non-top QCD multijet and $\PW(\cPZ)$+jets backgrounds. These triggers have been shown to have efficiencies of 92.2 $\pm$ 0.2\% and 98.2 $\pm$ 0.1\% for the muon and electron channels, respectively, based on independent studies of $\cPZ$ boson decays into $\Pgmp\Pgmm$ and $\Pep\Pem$ pairs. No azimuthal nor polar angle dependence is observed.

Muons are reconstructed using the information from the muon chambers and the silicon tracker and required to be consistent with the reconstructed primary vertex \cite{CMS_mu}. A kinematic selection requiring $\PT > 20$\gev and $|\eta| < 2.1$ is then used to select muon tracks for further analysis. Electrons are reconstructed using a combination of shower shape information and track/electromagnetic-cluster matching \cite{CMS_e}. A veto is applied to reject the electrons coming from photon conversions. To be retained for further analysis, electron candidates are required to have a transverse energy, $\ET > 30$\gev and $|\eta| < 2.5$, excluding the transition region between the barrel and endcap calorimeters, $1.44 < |\eta_\mathrm{c}| < 1.57$, where $\eta_\mathrm{c}$ is the pseudorapidity of the electromagnetic cluster.

Signal events are required to have only one isolated lepton, whose origin is consistent with the reconstructed $\Pp\Pp$ interaction vertex \cite{CMS_vertex}. Since the muon (electron) from a leptonic $\PW$ decay is expected to be isolated  from other high-$\PT$ particles in the event, its track is required to be isolated from other activity in the event. This is done by requiring a relative isolation ($I_{\mathrm{rel}}$) less than 0.05 for muons and 0.10 for electrons. A looser cut is used in the electron case to allow for the increased amount of radiation close to the track. Relative isolation is defined as $I_{\mathrm{rel}} = (I_\mathrm{charged} + I_\mathrm{neutral} + I_\mathrm{photon}) / \PT$, where $\PT$ is the transverse momentum of the lepton, and $I_\mathrm{charged}$, $I_\mathrm{neutral}$, and $I_\mathrm{photon}$ are the sums of the transverse energies of the charged and neutral hadrons and the photons reconstructed in a cone of $\Delta R < 0.3$ around the lepton direction, where  $\Delta R=\sqrt{(\Delta\phi)^2+(\Delta\eta)^2}$. The energy deposited by the lepton is explicitly removed from the sums by defining an exclusion cone of $\Delta R < 0.15$ around the lepton direction.  From studies based on $\cPZ$ decays in the data, the combined identification and isolation efficiencies for these selections are 83 $\pm$ 1\% for muons and 75 $\pm$ 1\% for electrons.

Semileptonic $\ttbar$ events have at least four hadronic jets (from the hadronization of the bottom and light quarks). Charged and neutral hadrons, photons, and leptons are reconstructed using the CMS particle-flow algorithm  \cite{ipartf} before they are clustered to form jets using the anti-$k_\mathrm{T}$ jet algorithm \cite{ktalg} with a cone size $\Delta R=0.5$. The jet clustering software used is \textsc{fastjet} version 2.4.2~\cite{fastjet1,fastjet2}.
At least one jet candidate with $\PT >$ 25 GeV and $|\eta| <  2.4$ is required, which must not overlap with a muon or electron candidate within $\Delta R < 0.3$.
Relative and absolute jet energy corrections~\cite{jec} are applied to the raw jet momenta to
establish a uniform jet response in $\PT$ and $\eta$ (with uncertainties of 3--6\%, dependent on
$\PT$ and $\eta$).
An offset correction, determined from simulation, is made to correct for the effects of event pileup (with an uncertainty $<$ 1.4\%).
There is an additional $\cPqb$-jet energy scale uncertainty which accounts for the differences in response between \textsc{Pythia} and \textsc{Herwig} simulations (${\sim} 3\%$). Lastly, a further 1.5\% uncertainty is added to allow for residual calibration differences during different run periods for a total of 4.7--7.0\%.

The neutrino from the leptonic $\PW$ decay escapes detection. Its presence is inferred  from a sizeable transverse energy imbalance in the detector.  The missing transverse energy (\ETslash) is defined as the negative of the vector sum of the transverse energies (\ET) of all of the particles found by the particle-flow algorithm. This is used as an event selection variable in both the muon and electron analyses to suppress the background from QCD multijet events, and is required to be greater than 20\GeV in both channels.

Because of the combined effects of the long $\cPqb$-quark lifetime ($\sim$1.5 ps) and the fact that the $\cPqb$ quarks are produced with a significant boost, the decays of $\cPqb$-flavored hadrons are quite different from the shorter lived hadronic states. Rather than having an origin consistent with the primary collision vertex, they can travel a measurable distance before decaying, resulting in a displaced decay vertex. The origin of the particles from the $\cPqb$ decay is thus typically inconsistent with the primary vertex position. In the case of a semileptonic $\cPqb$-hadron decay, this results in the production of a lepton with a displaced origin, that is also embedded inside a jet.
These characteristics can be used to identify $\cPqb$-quark jets and distinguish them from their non-$\cPqb$ counterparts. Here we use a displaced secondary vertex algorithm to tag $\cPqb$ decays
and suppress the background from $\PW(\cPZ)$+jet and QCD multijet events. The algorithm is described in detail in \cite{CMS_btag} and has a $\cPqb$-tag efficiency of 55\% with a light parton ($\cPqu,\cPqd,\cPqs,\cPg$) mistag rate of 1.5\% for jets with $\PT >$ 30 GeV in simulated QCD events. The event selection requires that at least one of the selected jets is $\cPqb$ tagged.

\section{Signal and Background Modeling}
\label{sec:sigBkgModel}

The efficiency for selecting lepton$+$jets signal events and the corresponding kinematic distributions are modeled using a simulated $\ttbar$ event sample. The simulation is performed using \textsc{MadGraph}~\cite{Alwall:2007st}, where the events containing top-quark pairs are generated accompanied by up to three extra partons in the matrix-element calculation. The parton configurations generated by \textsc{MadGraph} are processed with \textsc{Pythia 6.4}~\cite{Sjostrand:2006za} to provide fragmentation of the generated partons. The shower matching is done using the Kt-MLM prescription~\cite{Alwall:2007st}. The generated events are then passed through the full CMS detector simulation based on \textsc{Geant4}~\cite{Geant4}.

The production of $\PW$($\cPZ$) + jets events, where the vector boson decays leptonically, has a similar signature and constitutes the main background. These are also simulated using \textsc{MadGraph}.
We use a dynamical mass scale ($Q^{2}$ scale) of  $(m_{\PW/\cPZ})^2 + (\sum \PT^\mathrm{jet})^2$ for the renormalization and factorization scales for both the $\PW$+jets and $\cPZ$+jets simulations, and these scales are varied by factors of 2.0 and 0.5 in systematic studies.

In addition to the Monte Carlo (MC) generation using \textsc{MadGraph}, QCD multijet samples were produced using \textsc{Pythia}.

The QCD predictions for the top-quark pair production cross section are $157^{+23}_{-24}\unit{pb}$ in NLO \cite{Top_x1} and 164$^{+10}_{-13}\unit{pb}$ \cite{Aliev:2010zk,Langenfeld:2009wd} or 163$^{+11}_{-10}\unit{pb}$ \cite{Kidonakis:2010dk} in approximate NNLO. In each case the quoted uncertainties include the renormalization and factorization scale uncertainties, the uncertainties from the choice of parton distribution functions (PDF)s, and the uncertainty of the strong coupling constant ($\alpha_S$). For the scale uncertainty we have
varied the renormalization and factorization scales by factors of 2 and 0.5 around the central choice of $(2m_\cPqt)^2 + (\sum \PT^\mathrm{jet})^2$ with $m_\cPqt$ = 172.5 GeV.  The PDF and $\alpha_S$ uncertainties were determined by following the results obtained by using the MSTW2008~\cite{mstw08}, CTEQ6.6~\cite{cteq_2010}, and NNPDF2.0~\cite{nnpdf} sets and combining the results using the PDF4LHC prescriptions~\cite{pdf4lhc}.

Single top quark production is described in terms of the cross sections in the $s$, $t$ and $\cPqt\PW$ channels. The largest cross section is in the $t$ channel where the predicted NLO cross section is $\sigma_{t} = 64.6^{+3.4}_{-3.2}\unit{pb}$ from \textsc{mcfm}~\cite{Top_x1,mcfm2,mcfm3,mcfm4}. The result is given for a scale of $(m_{t})^2 + (\sum \PT^\mathrm{jet})^2$ and an uncertainty that is defined in the same way as for top-quark pair production. Similarly, the cross sections for the $\cPqt\PW$ and $s$ channels are predicted to be $\sigma_{\cPqt\PW}=10.6\pm0.8\unit{pb}$ and $\sigma_{s}=4.2\pm0.2\unit{pb}$~\cite{mcfm2}, respectively. A measurement of the $t$-channel cross section has been performed by CMS and is found to be consistent with the value predicted by the standard model \cite{CMS_stop}.

The inclusive NNLO cross section of the production of $\PW$ bosons decaying into leptons has been determined as $\sigma_{\PW\rightarrow l\nu} = 31.3 \pm 1.6\unit{nb}$ using \textsc{fewz}~\cite{fewz} (corresponding to a $k$-factor of 1.3), setting renormalization and factorization scales to $m_{\PW}^2$ with $m_\PW=80.398\GeV$. The uncertainty was determined in a similar way as for top-quark pair production. Finally, the Drell--Yan production cross section at NNLO has been calculated using \textsc{fewz} as $\sigma_{\cPZ/\gamma^*\rightarrow ll} (m_{ll}>50\GeV) = 3.0 \pm 0.1\unit{nb}$ where the scales were set to $m_{\cPZ}^2$ with $m_\cPZ=91.1876$\GeV.

The cross sections discussed above are used to normalize the simulated event samples.
The profile likelihood fit yields corrections to these values.
The normalizations for the final comparisons with the data are determined from fits to the data in control samples.

\section{Cross Section Measurements}

The following subsections discuss the measurement procedure and the results obtained from the analysis of the muon, electron and combined channels.

\subsection{Fit Procedure}

\label{sec:shyft_intro}

\begin{figure*}[htbp]
\centering
\includegraphics[angle=90,width=0.6\linewidth]{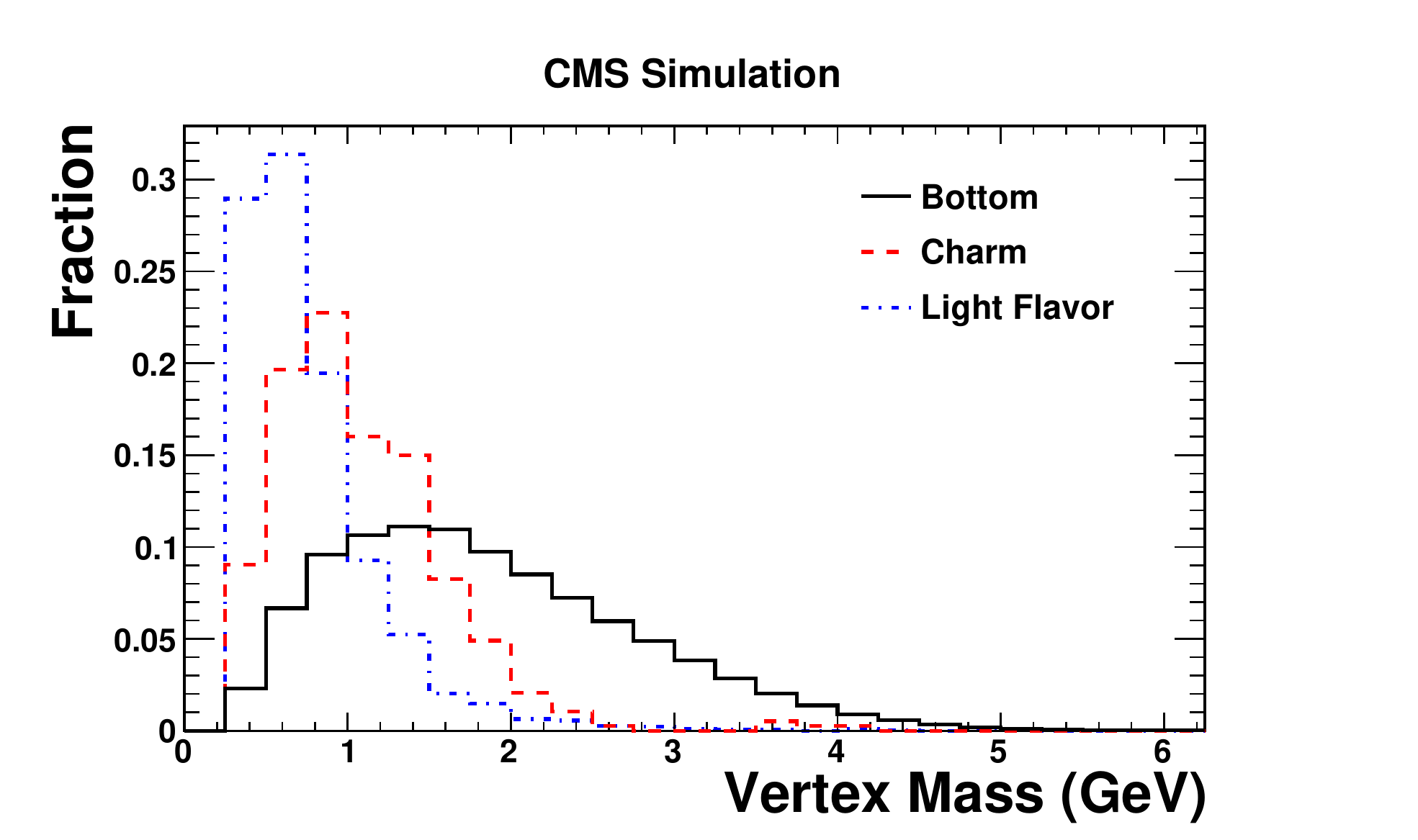}
  \caption{Secondary vertex mass distribution for bottom, charm, and light
    flavor jets. The bottom and charm templates are taken from simulated
    $\ttbar$ events and the light flavor shape is taken
    from simulated $\PW+$jet events.}
  \label{fig:secvtxMassTemplate}
\end{figure*}

To extract the $\ttbar$ cross section we perform a maximum likelihood fit to the
number of reconstructed jets ($j = 1$--4, ${\ge}5$), the number of $\cPqb$-tagged jets ($i = 1$, ${\ge2}$), and the secondary vertex mass distribution in the data. The secondary vertex mass is defined as the mass of the sum of the four-vectors of the tracks associated to the secondary vertex, assuming that each particle has the pion mass. It gives a good discrimination between the contributions from light and heavy flavor quark production (Fig.~\ref{fig:secvtxMassTemplate}).
We fit the data to the sum of signal and background shapes using a binned Poisson likelihood.

The templates for the fit are normalized to the expected event yields for 36\pbinv. The expected yield for each component in a given jet-multiplicity bin $j$ and tag-multiplicity bin $i$ is a function of the cross section, the jet energy scale (JES), the $\cPqb$-tag efficiency, and the background normalization. The JES and the $Q^2$-scales affect the jet multiplicity, and the $\cPqb$-tag efficiency impacts the number of $\cPqb$ tags. The background model depends on each of these and has additional contributions from the normalization and extrapolation from the sidebands in the data. Because these effects are expected to cause the largest uncertainties on the $\ttbar$ cross section measurement, and they are correlated, they are treated as nuisance parameters in the profile likelihood fit. The minimization of the likelihood thus provides simultaneous measurements of each contribution and the $\ttbar$ cross section.

We determine the background normalizations from data sideband regions and extrapolate to the signal region using various models. In particular, the four major backgrounds are $\PW+$jets, $\cPZ+$jets, single top quark, and QCD multijet production.

The templates for the $\PW$ and $\cPZ$ contributions are normalized such that the NNLO predictions are equal to unity. During the profile likelihood maximization,
the normalizations of these components are extracted. A full detector unfolding is not done, so this is not a meaningful measurement of the $W(Z)$+jets cross section, however the impact of the renormalization and factorization scales on the $\ttbar$ cross section is found to be smaller than that predicted by an ad hoc variation of the scales.

The $\PW$ and $Z+$ jets backgrounds come from $V$+$\cPqb$ jets, $V$+$\cPqc$ jets, and $V$+light flavor events.
The same $k$-factor of 1.30 (defined in Section~\ref{sec:sigBkgModel}) is applied to all three
flavors as inputs to the likelihood fit (although the three components are allowed to float independently
in the fit). An additional electroweak background is single-top ($s$ and $t$ and $tW$ channels) events.

The shape of the jet multiplicity distribution ($N_\mathrm{jet}$) depends on the choice of the jet $\PT$ threshold, and thus is also sensitive to the jet energy scale (JES).  In this sense, the fit is intrinsically able to determine the JES from the variations of the $N_{jet}$ distribution as a function of JES. The uncertainty in the $\cPqb$-tag efficiency is also extracted directly from the fit, by using the changes in the relative rates of 1-tag, and 2-tag events.  A larger $\cPqb$-tag efficiency will result in events moving from 1-tag to 2-tag samples.  In contrast, an overall increase in all tag bins together would indicate an increase in the $\ttbar$ cross section. The combined in-situ measurement of the yields of principal backgrounds and parameters describing main systematic uncertainties leads to a significant improvement over analyses which use more conventional techniques for tagged cross section measurements.

There are several ``nonprompt-$\PW$'' or ``QCD'' backgrounds for the muon and electron analyses.
The QCD background in the muon-plus-jets channel comes from multijet events with heavy flavor decays, kaon and pion decays in flight, and hadronic punch-through in the muon system. Because these are difficult to calculate to the required precision we derive these backgrounds from the data. The normalization is determined by using a comparison of data and simulation in the data sideband region with  $\ETslash < 20\GeV$. The ratio is used to scale the predicted yields for  $\ETslash > 20\GeV$.
The shapes and normalizations of the $\ttbar$, $\PW$ and $\cPZ$+jets are well described by the Monte Carlo simulation and are modeled that way.  The shape of the QCD component is derived from the nonisolated ($I_\mathrm{rel} > 0.2$) data.  Because of the correlations between \ETslash and isolation, the templates for the QCD estimate from the data that are taken from nonisolated samples are modified using the shape taken from the QCD simulation. This treatment is similar in spirit to the QCD treatment in the recent CMS $\PW$ and $\cPZ$ cross-section measurements \cite{CMS_WZ}.
The QCD rate in each jet bin is constrained to the average of the true \ETslash distribution of the nonisolated region, and the ``modified'' \ETslash distribution after accounting for correlations from the Monte Carlo simulation.
The QCD component in the fit is constrained to 100 percent of the rate or
half the difference between the results, whichever is greater.

The secondary vertex mass shapes for the fit are taken from the nonisolated data. Because of limited statistics, the $\ge 3$ jet sample is included as a single template, with separate normalizations for each jet bin.

Similarly, the electron-plus-jets channel is contaminated by photon conversions, jets with a high electromagnetic fraction, and heavy flavor decays. The normalization of this background is also estimated from a fit to the $\ETslash$ spectrum. However, the nonisolated sidebands do not accurately represent the shape of the $\ETslash$ distribution, and so the shape is determined by reversing at least two out of six of the electron identification criteria. The ensemble of these ``marginal failures'' makes a good representation of the shape of the $\ETslash$ background for this background source. This shape is then fit with the same $\ETslash$ procedure as in the muon case.

There are alternative control samples for both the jet energy scale, and the $\cPqb$-tag efficiency. The jet energy scale is measured as described in Section~\ref{sec:EventSelection}. The uncertainty measured there is used in a Gaussian constraint on the jet energy scale in the likelihood (approximately 4\%). The $\cPqb$-tagging efficiency and light quark mistag rate have been measured in an independent sample of QCD dijet events \cite{CMS_btag2}, and these values are input as Gaussian constraints on the parameters in the likelihood. Specifically, the $\cPqb$-tag efficiency scale factor is constrained to $1.0 \pm 0.2$, and the mistag rate scale factor is constrained to $1.0 \pm 0.1$. The technical implementation of these efficiencies in the likelihood is to weight the tagged jets in the simulation up or down by the data-to-simulation scale factor, and weight untagged jets in the simulation with zero weight.

The number of predicted events with respect to each contribution is given by Eqns.~(\ref{eqn:templatestop}--\ref{eqn:templateswqq}), for the $\ttbar$ signal and two of the $W+$jets backgrounds ($\PW$+$\cPqb$-jets and $\PW$+light flavor). There are similar terms for the other $\PW$+jets backgrounds, the single top, and the QCD multijet production. Thus we have

\begin{align}
\begin{split}
N_{\cPqt\cPaqt}^{\mathrm{pred}}(i,j) &= K_{\cPqt\cPaqt} \cdot N_{\cPqt\cPaqt}^{\mathrm{MC}}(i,j)  \cdot \\&
 P^{\cPqb~\mathrm{tag}}(i,j,R_{\cPqb~\mathrm{tag}}) \cdot 
 P^{\rm{mistag}}(i,j,R_{\mathrm{mistag}}) \cdot P^{\mathrm{JES}}(i,j,R_{\mathrm{JES}})\label{eqn:templatestop}\\
\end{split}\\
\begin{split}
N_{\PW\cPqb\cPqb}^{\mathrm{pred}}(i,j) &= K_{\PW\cPqb\cPaqb} \cdot N_{\PW\cPqb\cPqb}^{\mathrm{MC}}(i,j )   \cdot\\&
  P^{\cPqb~\mathrm{tag}}(i,j,R_{\cPqb~\mathrm{tag}}) \cdot 
   P^{\mathrm{mistag}}(i,j,R_{\mathrm{mistag}}) \cdot P^{\mathrm{JES}}(i,j,R_{\mathrm{JES}}) \cdot P^{Q^2}(i,j,R_{Q^2}) \label{eqn:templateswbb}\\
\end{split}\\
\begin{split}
N_{\PW\cPq\cPq}^{\mathrm{pred}}(i,j) &= K_{W\cPq\cPaq} \cdot N_{\PW\cPq\cPq}^{\mathrm{MC}}(i,j)    \cdot\\&
  P^{\mathrm{mistag}}(i,j,R_{\mathrm{mistag}}) \cdot 
  P^{\mathrm{JES}}(i,j,R_{\mathrm{JES}}) \cdot P^{Q^2}(i,j,R_{Q^2}) \label{eqn:templateswqq}\\
\end{split}
\end{align}

where $K_{\cPqt\cPaqt}$ is the fitted scale factor for the NLO prediction for $\ttbar$; $i$ and $j$ run over tags and jets, respectively; $K_{\PW\cPqb\cPaqb}$ is the fitted scale factor for the NNLO prediction for $\PW\cPqb\cPaqb$ (etc); $N_x^{\rm{MC}}(i,j)$ is the number of events expected for sample $X$, derived from Monte Carlo and corrected with data-to-Monte-Carlo scale factors. The $P^X(i,j,R_X)$ factors are multiplicative functions accounting for the relative differences with respect to the input expected yield, as a function of the assumed value $R_X$ of nuisance parameter $X$ (i.e., $\cPqb$-tag efficiency, jet energy scale, etc). These are interpolated from  various configurations in the simulation with polynomials. The convention chosen is that the nominal event yield is at $R_X = 0$ (i.e., no variation in parameter $X$), and $P(i,j,R_X) = 1.0$ (i.e., multiplicative factor of 1.0 by default). The ``$+1\sigma$'' variation is at $R_X=1$, and the ``$-1\sigma$'' variation is at $R_X=-1$.

The fit minimizes the negative log likelihood, summing over the histogram bins ($k$) of the secondary vertex mass, the number of jets ($j$), and the number of tags ($i$). The various constraints (described above) are included as Gaussian penalty terms on the variables, which are represented by $C_X$. The full profile likelihood expression is

\begin{equation}\begin{split}
  -2 \ln{L} =& -2 \Bigg\{\sum_{i,j}^\text{tag,jet}\sum_k^\text{bins}\left(\ln{
  \mathcal{P}(N^\text{obs}_k (i,j), N^\text{exp}_k (i,j))}\right) - 
  \frac{1}{2}\sum_l^\text{constraints} \frac{(C_X - \hat{C}_X)^2} {\sigma_{C_X}^2}\Bigg\}
  \label{eqn:negloglike}
\end{split}\end{equation}

where $\mathcal{P}$ is a Poisson probability that the predicted yield ($N^\text{exp}$) given by the various components statistically overlaps with the data ($N^{\rm{obs}}$) in each tag/jet bin $i,j$, given by

\begin{equation}
  \ln{\mathcal{P}(x, y)} = x \ln{y} - y - \ln{\Gamma (x + 1)}
  \label{eqn:poisson}
\end{equation}

where $\Gamma(x)$ is the Gamma function.

Table~\ref{table:shyft_constraints} shows a summary of all of the inputs to the profile likelihood,
as well as the constraints.

\begin{table*}[htbp]
\begin{center}
\caption{\label{table:shyft_constraints} {Inputs to the profile likelihood, along with constraints.}}
\begin{tabular}{| l | c |}
\hline
  \multicolumn{1}{|c|}{$\rm{Quantity}$}  & Constraint (\%)  \\
\hline
$\cPqb$-tag Efficiency Scale Factor &  20 \\
$\cPqb$-tag Mistag Scale Factor     &  10 \\
Jet energy scale relative to nominal&  4  \\
$\PW+$jets renormalization/factorization scales & $^{+100}_{-50}$ \\
$\PW+$jets background normalization   &  unconstrained \\
QCD background normalization        &  100 \\
Single-top background normalization &  30 \\
$\cPZ+$jets background normalization   &  30 \\
\hline
\end{tabular}
\end{center}
\end{table*}

There are also a number of systematic uncertainties that are not included directly in the profile likelihood and hence are taken as additional systematic uncertainties outside of the fit result. The largest of these is the systematic uncertainty due to the overall luminosity determination.  It has also been shown on independent samples of $\cPZ$ $\to$ $\Pe\Pe$ and $\cPZ$ $\to$ $\Pgm\Pgm$ events, that the efficiencies for triggering, reconstructing, and identifying isolated leptons of this type are very similar in the data and simulations. We have corrected for the small differences observed. The effect of these uncertainties are not included in the profile likelihood, and hence are taken as an additional systematic uncertainty of 3\%.

There are a number of theoretical uncertainties in the signal modeling that are not included in the profile likelihood at this time. They include differences in the $\ttbar$ signal due to renormalization and factorization scales, the amount of initial and final state radiation present, the parton distribution function model, and the matching scale for the matrix-element to parton-shower matching scheme. These are computed from dedicated simulated samples by varying the theoretical parameters of interest according to conservative variations around the reference value. The exception is the parton distribution functions, which are varied by reweighting the sample according to variations in the underlying parton distribution function parameterizations~\cite{pdf4lhc}. The numerical impact of each of these is taken as a systematic uncertainty. Specifically these are 2\% for $\ttbar$ $Q^2$ modeling; 2\% for initial and final state radiation modeling; 1\% for the matrix-element to parton-shower matching in the $\ttbar$; and 3\% for the parton distribution function differences.

In all of the cases that are described below, the robustness of the statistical procedure is demonstrated with a priori pseudo-experiments where the expected yields and the parameters in the profile likelihood are sampled randomly according to Poisson or Gaussian statistics (as appropriate). In cases where the true frequentist statistical coverage is not achieved due to the limitations of the profile likelihood method, coverage is assured by correcting for the slight biases (of order 1--2\% in the central values and/or uncertainties).

\subsection{Muon + Jet Analysis}
\label{sec:shyft_mu}

For the muon channel we make two modifications to the basic event selection discussed above. Instead of using the isolation cut of $I_\mathrm{rel}<$ 0.05, we require that the selected muon tracks are found to be isolated by the particle-flow reconstruction and pass a cut of $I_\mathrm{rel}<$ 0.15.  The results of the fit are shown in Table~\ref{table:muon_fit_12}. The fit considers events with one tag (1-tag) and two and more tags (2-tag) separately, giving  nine jet-tag `bins' (subsamples) which are fit by the joint likelihood.  Table~\ref{table:muon_fit_12} lists the observed and fitted rates for each jet-tag bin.

\begin{table*}[htbp]
\begin{center}
\caption{\label{table:muon_fit_12} {Results of the fit for muon+jets events with at least 1 $\cPqb$ tag.}}
\begin{tabular}{| l | r | r | r | r | r | r | r | r | r |}
\hline
               &       Data  &  \multicolumn{1}{c|}{$\rm{Fit}$}  &   \multicolumn{1}{c|}{$\ttbar$}  &  $\cPqt(\cPaqt)$
               &   \multicolumn{1}{c|}{\PW}  &  \multicolumn{1}{c|}{\PW}  &   \multicolumn{1}{c|}{\PW}  &  \multicolumn{1}{c|}{\cPZ}  &        QCD  \\
               &    &   &   &   & +$\cPqb$ jets & +$\cPqc$ jets & +$\cPq\cPaq$ & +jets &      \\
\hline
1 jet  1 $\cPqb$ tag   &         505   &       504.0   &        13.3   &        25.0   &        94.2   &       255.1   &        81.9   &        13.9   &        20.6   \\
2 jets 1 $\cPqb$ tag   &         314   &       318.2   &        51.0   &        29.4   &        82.6   &        97.7   &        35.0   &         7.3   &        15.1   \\
3 jets 1 $\cPqb$ tag   &         166   &       158.5   &        78.3   &        14.8   &        29.5   &        21.9   &        10.4   &         2.8   &         0.8   \\
4 jets 1 $\cPqb$ tag   &          85   &        89.2   &        60.6   &         4.9   &        12.8   &         5.5   &         3.3   &         1.3   &         0.8   \\
$\ge$5 jets 1 $\cPqb$ tag   &          45   &        43.8   &        34.6   &         1.5   &         4.6   &         1.8   &         0.9   &         0.2   &         0.2   \\
2 jets $\ge$2 $\cPqb$ tags  &          29   &        24.1   &        14.7   &         3.3   &         5.3   &         0.5   &         0.2   &         0.2   &         0.0   \\
3 jets $\ge$2 $\cPqb$ tags  &          37   &        44.0   &        35.2   &         3.8   &         3.9   &         0.8   &         0.0   &         0.2   &         0.0   \\
4 jets $\ge$2 $\cPqb$ tags  &          41   &        41.0   &        36.2   &         1.9   &         2.4   &         0.4   &         0.0   &         0.1   &         0.0   \\
$\ge$5 jets $\ge$2 $\cPqb$ tags  &          27   &        26.0   &        24.0   &         0.8   &         1.0   &         0.1   &         0.1   &         0.1   &         0.0   \\
\hline
Total          &        1249   &      1248.8   &       347.8   &        85.4   &       236.2   &       383.8   &       131.8   &        26.1   &        37.6  \\
\hline
\end{tabular}
\end{center}
\end{table*}

Table~\ref{table:SHYFT_Mu_systematics} lists the systematic uncertainties from the fit. These include both the theoretical uncertainties from the $\ttbar$ modeling and the corrections used to match the simulations to the data and give a total uncertainty of 4.3\% for the $\ttbar$ signal model. The unclustered energy in the detector results in an additional resolution uncertainty of $<$1\% on the \ETslash scale. We combine these with the data-simulation uncertainties due to the jet energy scale and jet resolution modeling, the $\cPqb$-tag efficiency and mistag rate and obtain a total systematic uncertainty of 12.5\%. For illustrative purposes, in Table~\ref{table:SHYFT_Mu_systematics}, we have broken up the pieces of the profile likelihood and quote the uncertainties due to the individual contributions. These are the result of fixing all of the other parameters of the likelihood and only allowing the chosen term to vary.

\begin{table*}[htbH]
\begin{center}
\caption{
  \label{table:SHYFT_Mu_systematics}
  \label{table:SHYFT_Elsys}
  \label{table:SHYFT_syssumm}
   List of systematic
   uncertainties for the muon + jet, electron + jet, and combined analyses.
   Due to the correlation
   between parameters in the fit, the combined number is not the sum of
   the squares of the contributions.}
\begin{tabular}{|c|c|c|c|}
\hline
Source & Muon      & Electron & Combined \\
              & Analysis & Analysis & Analysis \\
\hline
 \multicolumn{1}{|c|}{Quantity}  &  \multicolumn{3}{c|}{Uncertainty~(\%)}  \\
\hline
Lepton ID/reco/trigger                         & \multicolumn{3}{c|}{$3$} \\
\ETslash resolution due to unclustered energy      & \multicolumn{3}{c|}{$< 1$} \\
$\ttbar$+jets $Q^2$ scale                      & \multicolumn{3}{c|}{$2$} \\
ISR/FSR                                        & \multicolumn{3}{c|}{$2$} \\
ME to PS matching                              & \multicolumn{3}{c|}{$2$} \\
PDF                                            & \multicolumn{3}{c|}{$3$} \\
\hline
\multicolumn{1}{|c|}{Profile Likelihood Parameter}  & \multicolumn{3}{c|}{Uncertainty~(\%)}  \\
\hline
Jet energy scale and resolution               & $10$ & $9$ & $7$\\
$\cPqb$-tag efficiency                        & $9$  & $8$ & $8$\\
$\PW$+jets $Q^2$ scale                          & $4$  & $3$ & $9$\\
\hline
Combined                                      & 13 & 12 & 12\\
\hline
\end{tabular}
\end{center}
\end{table*}

This yields a cross section measurement of

\begin{equation}
\sigma_{\ttbar} =  145   \pm 12  ~\mathrm{(stat.)} \pm 18  ~\mathrm{(syst.)} \pm 6  ~\mathrm{(lumi.)}\unit{pb,}
\end{equation}

where the last uncertainty corresponds to the 4\% uncertainty on the total integrated luminosity \cite{CMS_Lumi}. The fit provides in-situ measurements of the scale factors for both $\cPqb$ tagging and the jet energy scale. We obtain a result of $98 \pm 6$\% for the $\cPqb$-tag scale factor and $92\pm 9$\% for the jet energy calibration. The scale factors for the $\PW$+$\cPqb$ jets and $\PW$+$\cPqc$ jets components indicate that the contributions in the data may be larger than what is expected by the predictions. For the $\PW$+$\cPqb$ jets component we find a cross section scale factor of 2.6 $^{+0.8}_{-0.7}$ and for the $\PW$+$\cPqc$ jets contribution we obtain 1.3 $^{+0.3}_{-0.2}$. It is also found that the $\PW$+jets data are slightly harder than the central value of the renormalization and factorization scales chosen.

\subsection{Electron + Jets Analysis}

\label{sec:shyft_ele}

The analysis in the electron channel is performed in the same way as
for the muon case. The results are shown in
Table~\ref{table:electron_fit_12}. The fit was performed in the same
manner as for the muon channel, resulting in nine jet-tag `bins'
(subsamples), which were fit by a joint likelihood, as described in
Section~\ref{sec:shyft_intro}.  Table~\ref{table:electron_fit_12}
lists the observed and fitted rates for each jet-tag bin. Note that the fit parameters are unbounded to avoid problems with bias. This can result in negative values for the event counts, as in the case of the 4 jets 1 tag yield for QCD in Table~\ref{table:electron_fit_12}.

\begin{table*}[hbtp]
\begin{center}
\caption{\label{table:electron_fit_12} {Results of the electron+jets fit for events with at least 1 $\cPqb$ tag.}}
\begin{tabular}{| l | r | r | r | r | r | r | r | r | r |}
\hline
               &       Data  &  \multicolumn{1}{c|}{$\rm{Fit}$}  &   \multicolumn{1}{c|}{$\ttbar$}  &  $\cPqt(\cPaqt)$
               &   \multicolumn{1}{c|}{W}  &  \multicolumn{1}{c|}{\PW}  &   \multicolumn{1}{c|}{\PW}  &  \multicolumn{1}{c|}{\cPZ}  &        QCD  \\
               &    &   &   &   & +$\cPqb$ jets & +$\cPqc$ jets & +$\cPq\cPaq$ & +jets &      \\
\hline
1 jet  1 $\cPqb$ tag   &         388   &       389.8   &         6.0   &        14.1   &        42.4   &       249.5   &        26.9   &         3.1   &        47.9   \\
2 jets 1 $\cPqb$ tag   &         252   &       245.3   &        31.7   &        21.0   &        44.4   &       104.0   &        14.8   &         3.3   &        26.2   \\
3 jets 1 $\cPqb$ tag   &         159   &       156.0   &        62.7   &        12.2   &        23.5   &        34.4   &         5.1   &         2.1   &        16.0   \\
4 jets 1 $\cPqb$ tag   &          71   &        80.7   &        60.6   &         4.8   &         8.2   &         9.2   &         1.4   &         0.8   &        $-4.3$   \\
$\ge$5 jets 1 $\cPqb$ tag   &          57   &        52.1   &        40.9   &         1.6   &         2.6   &         3.0   &         0.5   &         0.4   &         3.0   \\
2 jets $\ge$2 $\cPqb$ tags  &          14   &        19.9   &         9.4   &         2.3   &         4.7   &         1.2   &         0.1   &         0.2   &         2.0   \\
3 jets $\ge$2 $\cPqb$ tags  &          39   &        38.1   &        29.1   &         3.1   &         3.8   &         0.9   &         0.0   &         0.2   &         0.9   \\
4 jets $\ge$2 $\cPqb$ tags  &          37   &        41.3   &        37.1   &         1.9   &         1.9   &         0.4   &         0.0   &         0.1   &         0.0   \\
$\ge$5 jets $\ge$2 $\cPqb$ tags  &          37   &        30.7   &        28.8   &         0.9   &         0.8   &         0.2   &         0.0   &         0.1   &         0.0   \\
\hline
Total          &        1054   &      1053.8   &       306.3   &        61.8   &       132.1   &       402.7   &        48.9   &        10.3   &        91.6  \\
\hline
\end{tabular}
\end{center}
\end{table*}

The resulting cross section is

\begin{equation}
\sigma_{\ttbar} =  158   \pm 14  ~\mathrm{(stat.)} \pm 19 ~ \mathrm{(syst.)} \pm 6  ~\mathrm{(lumi.)}\unit{pb.}
\end{equation}

From the fit we obtain a result of $97\pm 6$\% for the $\cPqb$-tag scale factor and $103\pm 8$\% for the jet energy calibration. The scale factors for the $\PW$+$\cPqb$ jets and $\PW$+$\cPqc$ jets components are 1.4 $^{+0.8}_{-0.6}$ and 1.4 $^{+0.4}_{-0.3}$, respectively. These are in agreement with the results from the muon channel. The contributions to the systematic uncertainty are summarized in Table~\ref {table:SHYFT_Elsys}.
\subsection{Simultaneous Muon and Electron Channel Analysis}
\label{sec:shyft_combined}

\begin{figure*}[htb]
\centering
    \includegraphics[angle=90,width=0.8\textwidth]{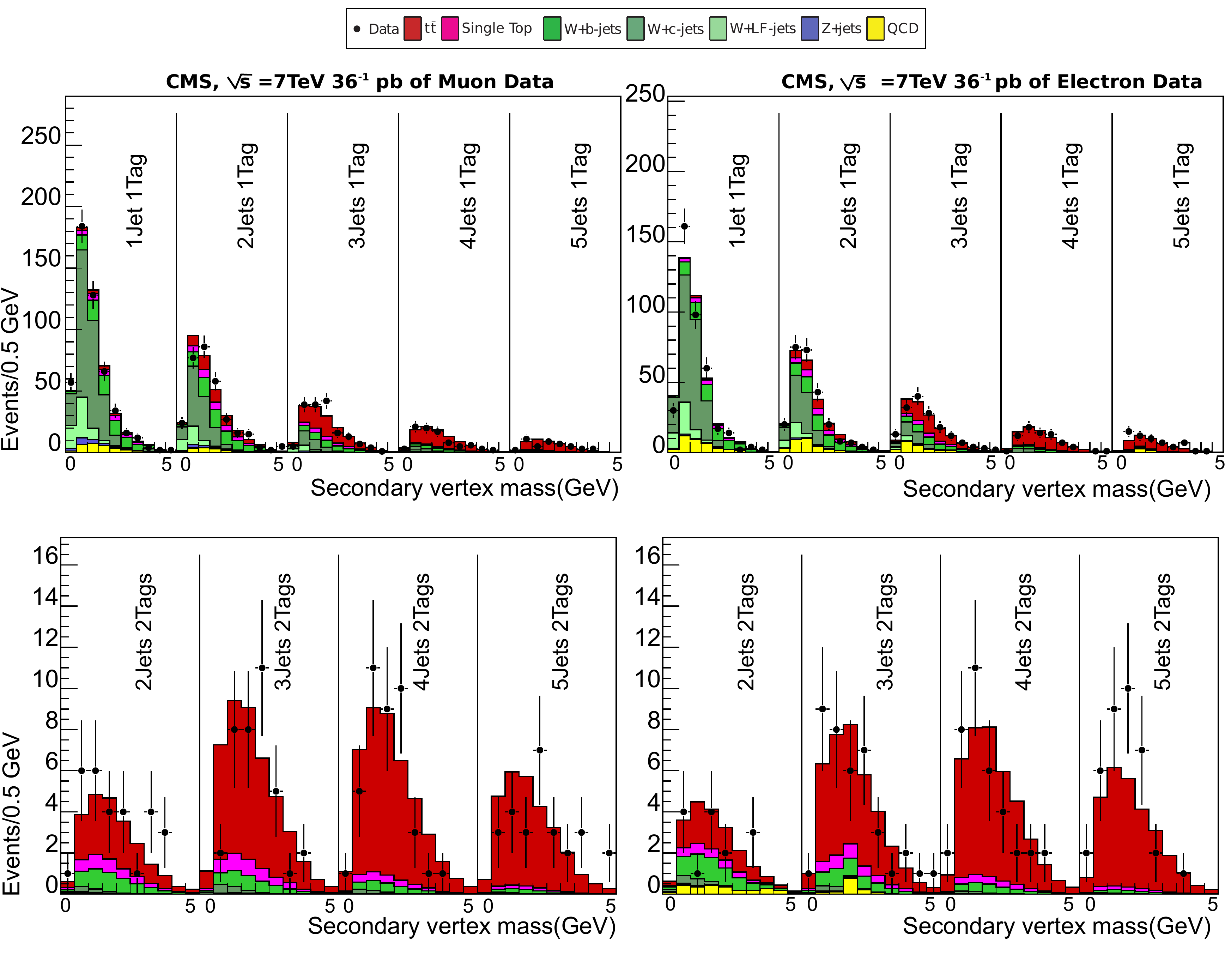}
  \caption[Results of the combined muon and electron fit.]{Results of
    the combined muon and electron channel fit. The muon channel is
    shown on the left and the electron channel on the right. The plots
    on the top are for exactly 1 $\cPqb$ tag and those on the bottom are
    for $\ge$2 $\cPqb$ tags. The histograms within the top panel correspond
    to events with 1, 2, 3, 4 and $\ge$5 jets, respectively, while the bottom panel shows histograms corresponding to events with 2, 3, 4 and  $\ge$5jets. }
  \label{fig:combined_tagged_35pb}
\end{figure*}

\begin{figure}[htb]
\centering
\begin{tabular}{cc}
   \includegraphics[angle=90,width=0.5\linewidth]{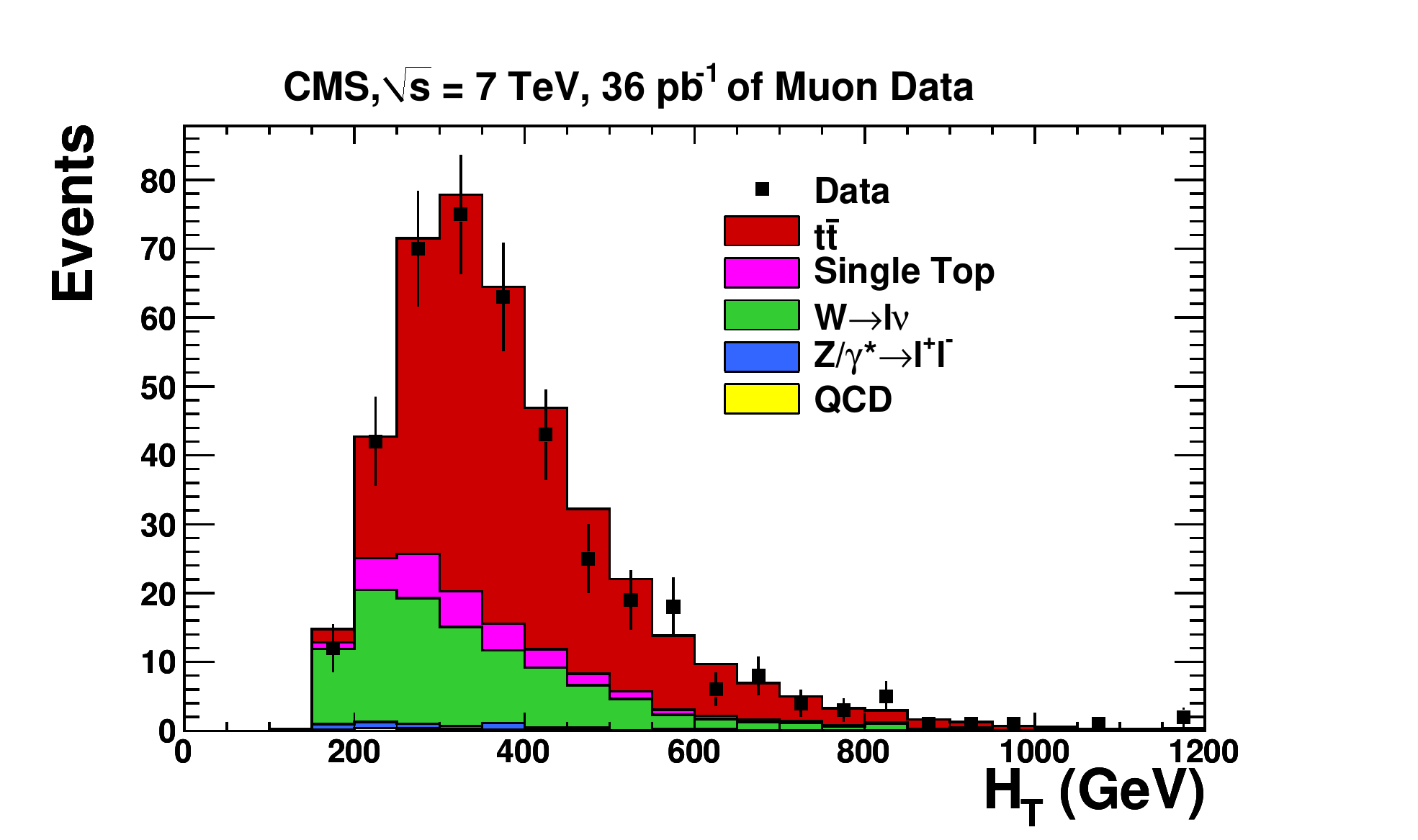} &
   \includegraphics[angle=90,width=0.5\linewidth]{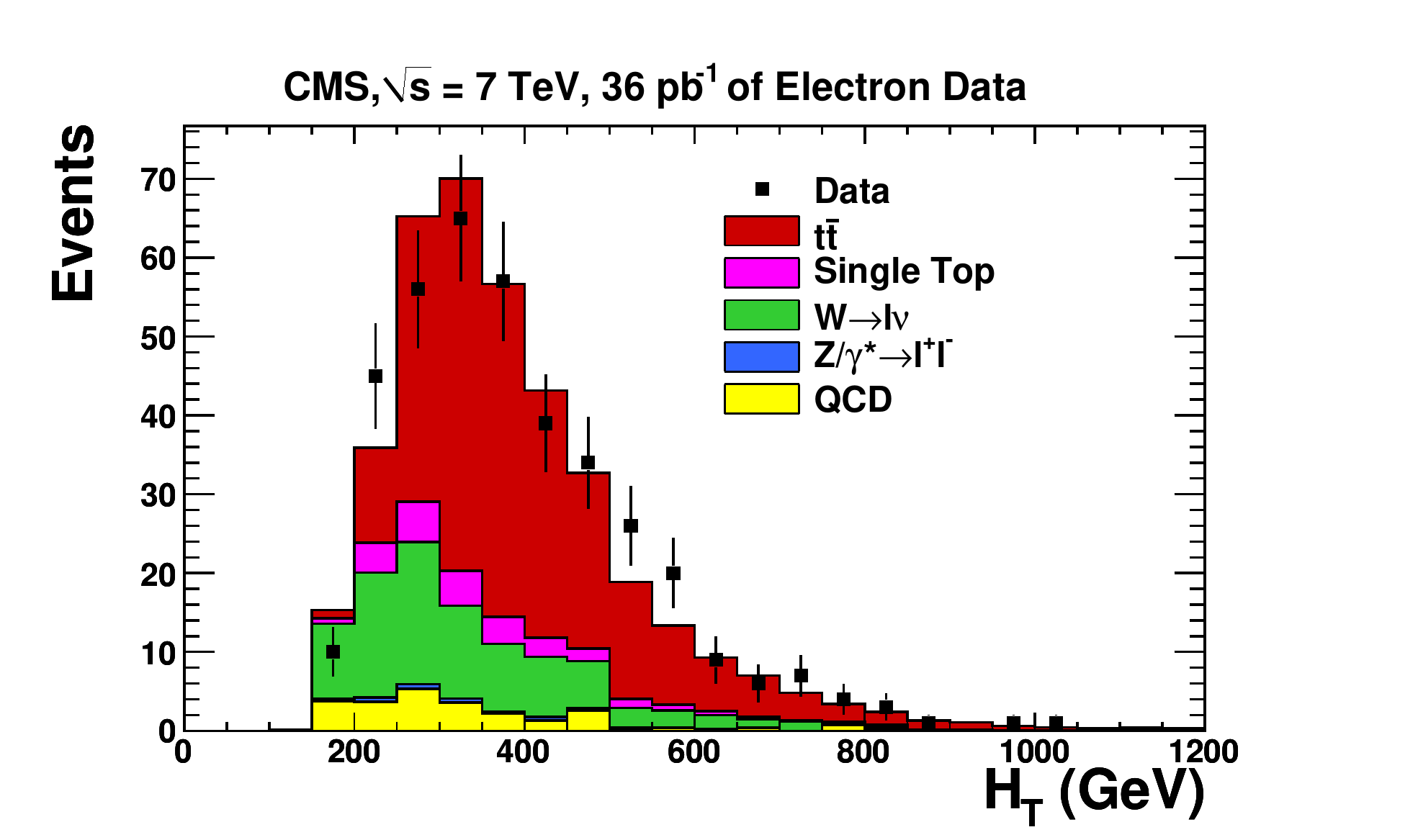} \\
   \includegraphics[angle=90,width=0.5\linewidth]{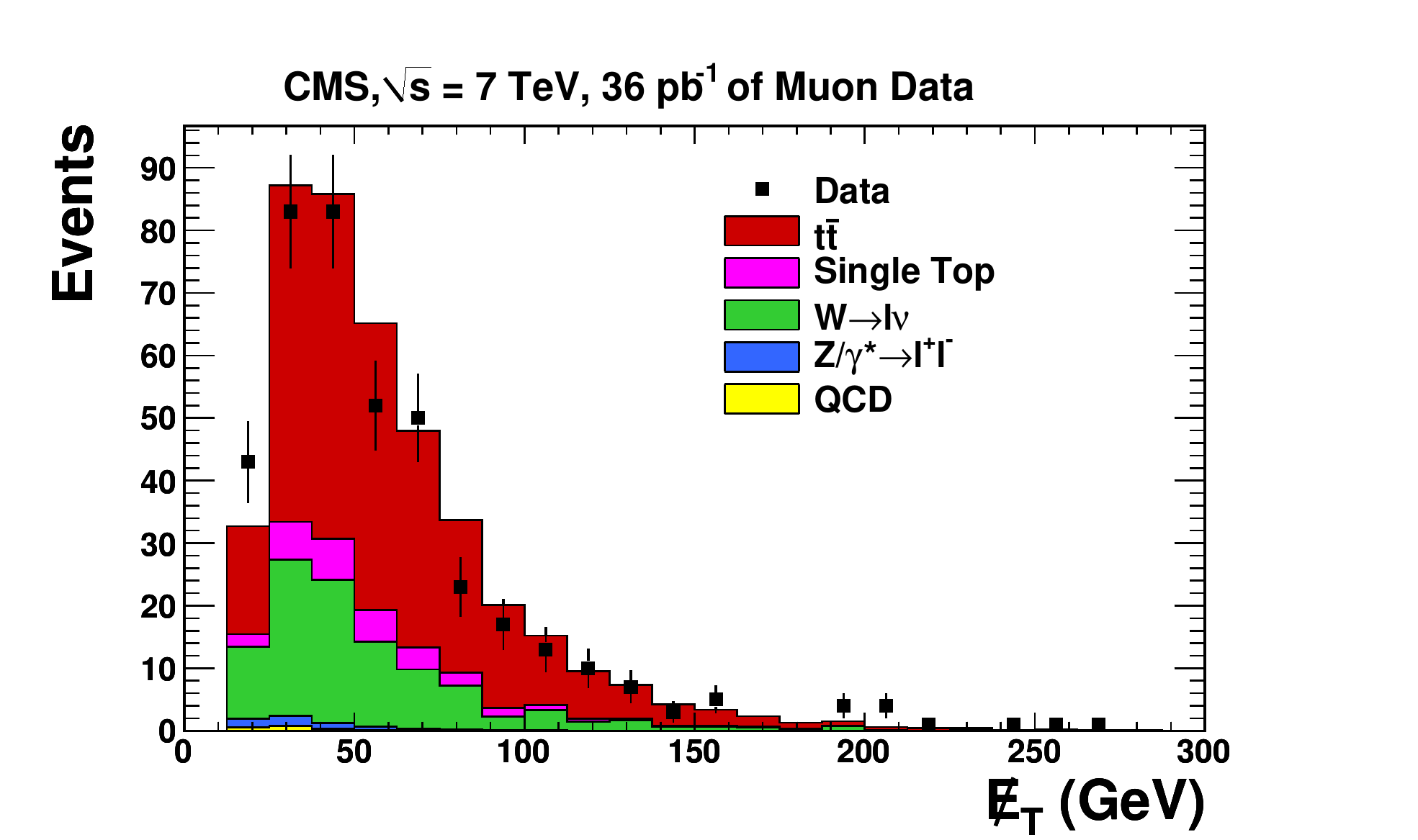} &
   \includegraphics[angle=90,width=0.5\linewidth]{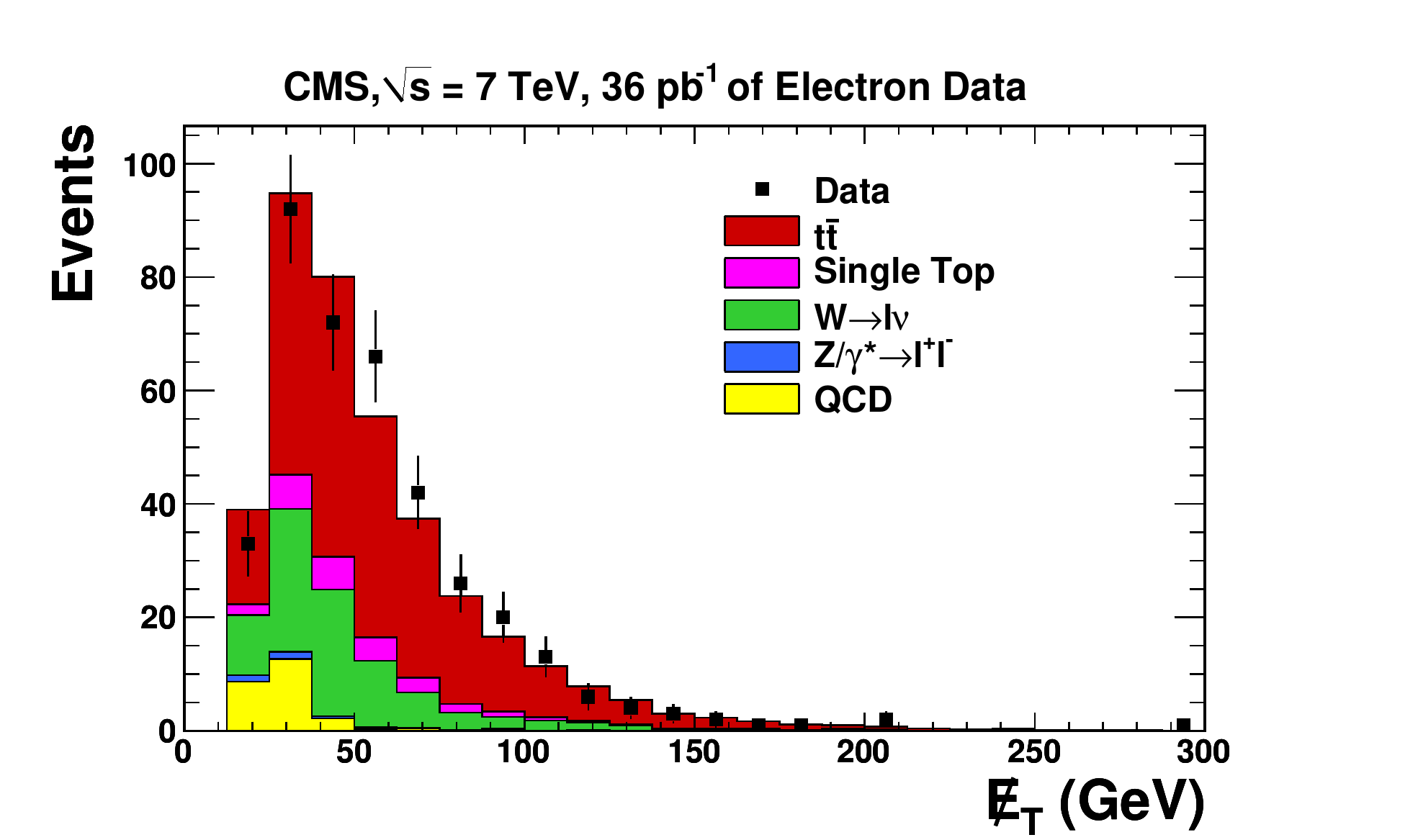} \\
   \includegraphics[angle=90,width=0.5\linewidth]{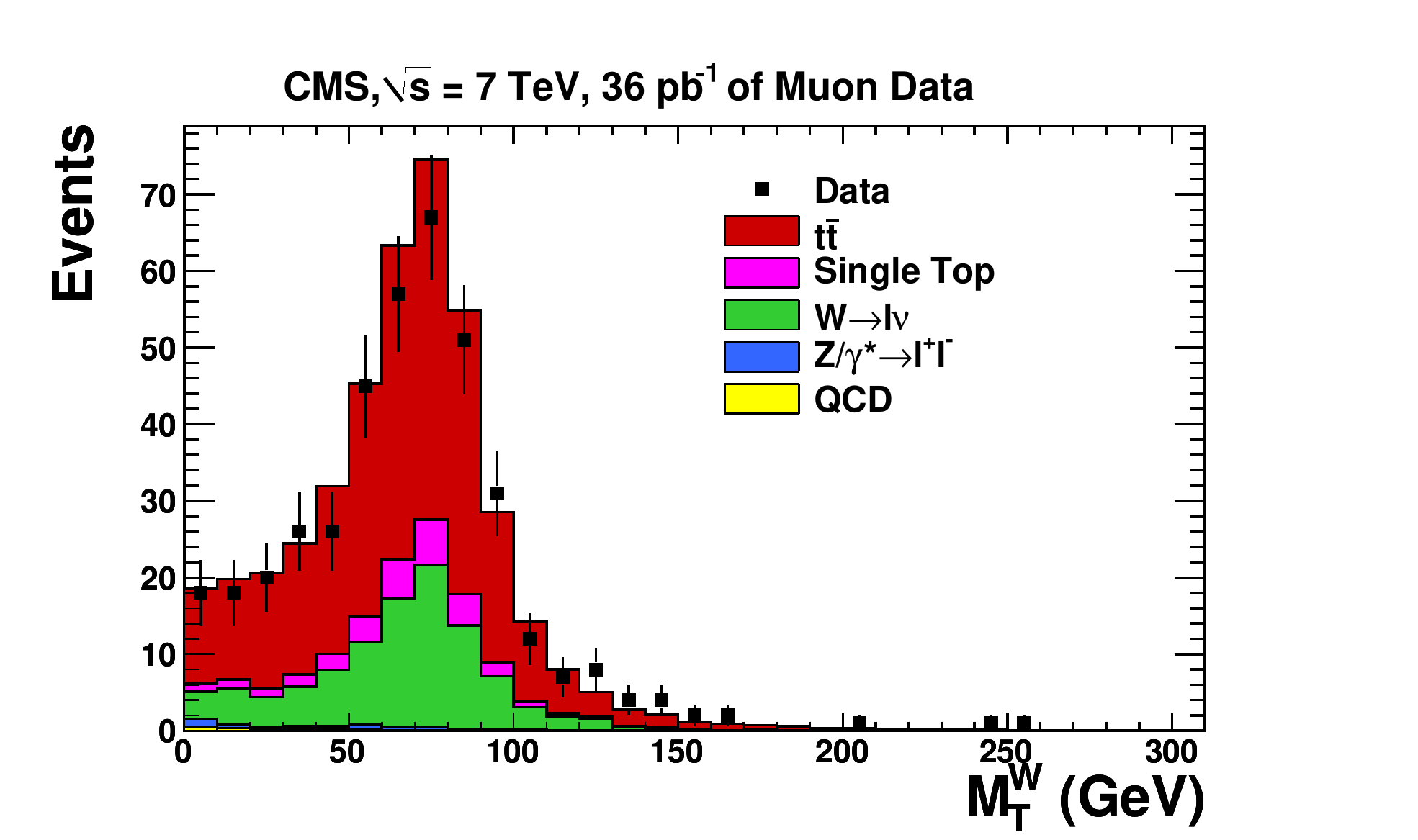} &
   \includegraphics[angle=90,width=0.5\linewidth]{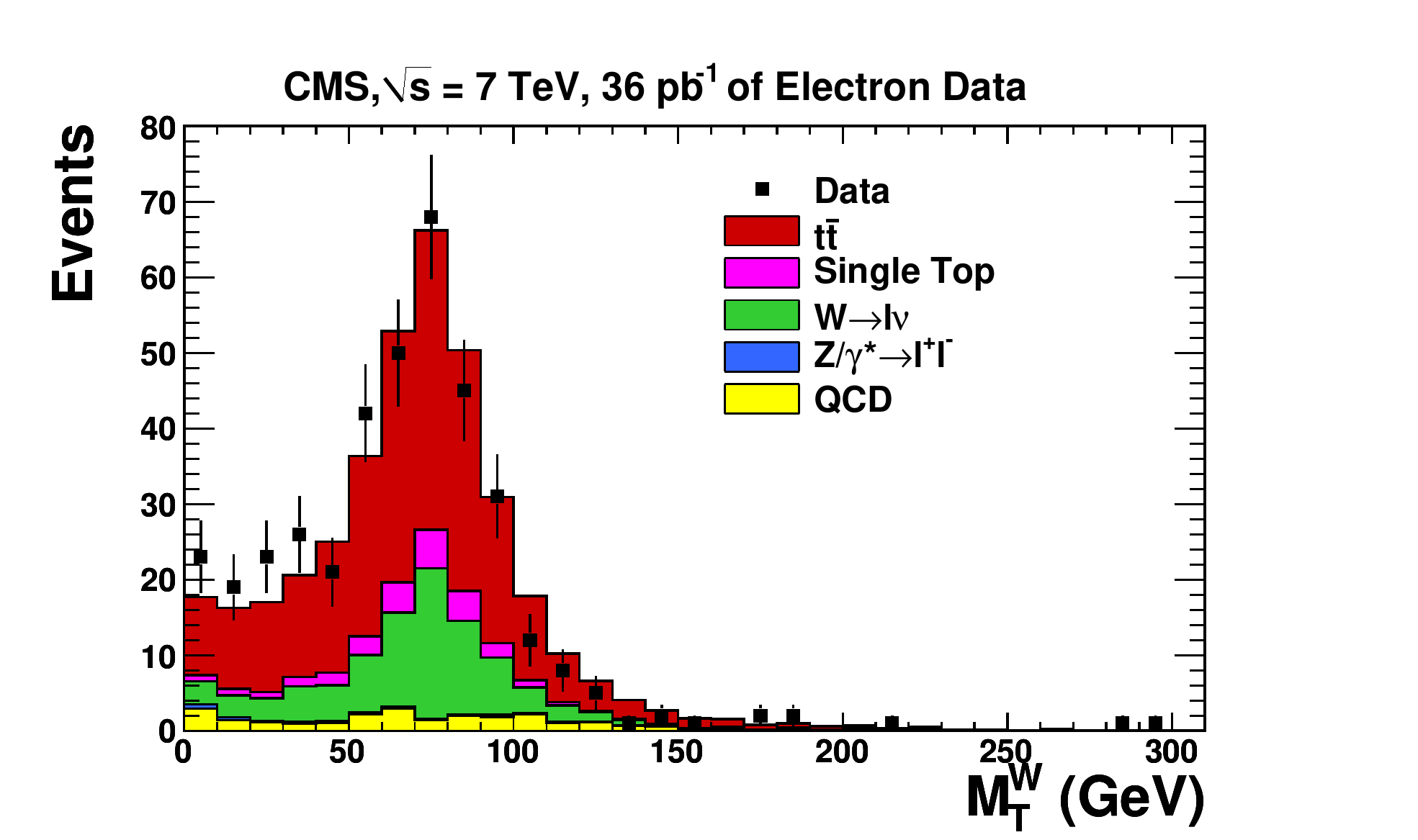} \\
\end{tabular}
  \caption[Kinematic distributions]{Kinematic distributions of the
    total transverse energy ($H_\mathrm{T}$), the missing transverse energy
    (\ETmiss)
    and the transverse mass of the $\PW$
    (\ETslash). The plots are for $\ge3$~jets and $\ge1\cPqb$~tag and the
    comparison histograms correspond to the fitted values. The muon channel
    is shown on the left and the electron channel is on the right. }
  \label{fig:combined_tagged_35pb_prettyplots}
\end{figure}

Having established the consistency of the separate channel measurements, we now proceed to perform a combined fit to both channels. To establish our best measurement, we repeat the fit procedure and apply it simultaneously to the data in both the electron and muon channels.
We find that the resulting fitted event yields in each tag category are in good agreement with those obtained from the separate channel fits (Tables~\ref{table:muon_fit_12} and \ref{table:electron_fit_12}). Figure~\ref{fig:combined_tagged_35pb} shows the comparison of the corresponding observed and fitted vertex mass distributions. Figure~\ref{fig:combined_tagged_35pb_prettyplots} shows the data for $\ge3$ jets and $\ge1~\cPqb$-tag, and the fit results for the total transverse energy of the event ($\HT$), the missing transverse energy ($\ETmiss$), and the transverse mass of the $\PW$ ($M_\mathrm{T}^\PW$).  We find good agreement in all cases.

\begin{table*}[hbtp]
\begin{center}
\caption{\label{table:corr_combined_shyfts_12}
{Correlation matrix of the fit to the combined electron and muon data samples.}}
\begin{tabular}{| l | rrrrrrrrrr|}
\hline
               & $\ttbar$ & $\cPqt(\cPaqt)$ & \multicolumn{1}{c}{\PW} & \multicolumn{1}{c}{\PW} & \multicolumn{1}{c}{\PW} &\multicolumn{1}{c}{\cPZ} & $Q^2$ & $\cPqb$ tag & JES & $R_{mistag}$ \\
              &    &    &  +$\cPqb$ jets & +$\cPqc$ jets & +$\cPq\cPaq$ & +jets &    &    &    &    \\
\hline

  $\ttbar$     &      1.0  &  $-0.1$  &   0.1  &   0.5  &   0.2  &   0.0  &   0.3  &  $-0.7$  &  $-0.6$  &  $0.0$  \\
$\cPqt(\cPaqt)$     &    $ -0.1$  &   1.0  &  $-0.3$  &   0.0  &   0.0  &  0.0  &   0.1  &  $0.1$  &   0.1  &   0.0  \\
  $\PW$+$\cPqb$ jets     &      0.1  &  $-0.3$  &   1.0  &   0.4  &   0.4  &  $ 0.0$  &   0.6  &  $-0.2$  &   0.0  &   0.0  \\
  $\PW$+$\cPqc$ jets     &      0.5  &   0.0  &   0.4  &   1.0  &  $-0.1$  &  0.0  &   0.6  &  $-0.5$  &  $-0.2$  &  $0.0$  \\
  $\PW$+$q\cPaq$ &      0.2  &   0.0  &   0.4  &  $-0.1$  &   1.0  &  $-0.1$  &   0.5  &  $-0.2$  &  $-0.2$  &  $-0.3$  \\
  $\cPZ+$jets     &      0.0  &   0.0  &  $0.0$  &   0.0  &  $-0.1$  &   1.0  &   0.0  &  $0.0$  & $0.0$  &   0.0  \\
     $Q^2$     &      0.3  &   0.1  &   0.6  &   0.6  &   0.5  &   0.0  &   1.0  &  $-0.2$  &   0.1  &   0.0  \\
   $\cPqb$ tag     &     $-0.7$  &  $-0.1$  &  $-0.2$  &  $-0.5$  &  $-0.2$  &  $0.0$  &  $-0.2$  &   1.0  &   0.3  &   0.0  \\
       JES     &     $-0.6$  &   0.1  &   0.0  &  $-0.2$  &  $-0.2$  &  $0.0$  &   0.1  &   0.3  &   1.0  &   0.0  \\
  $R_{mistag}$   &     $0.0$  &   0.0  &   0.0  &  $0.0$  &  $-0.3$  &   0.0  &   0.0  &   0.0  &   0.0  &   1.0  \\
\hline
\end{tabular}
\end{center}
\end{table*}

The correlation matrix for the combined fit is listed in Table~\ref{table:corr_combined_shyfts_12}. All of the terms are as defined in the text.
The combined analysis cross section measurement is

\begin{equation}
\sigma_{\ttbar} =  150   \pm 9 ~ \mathrm{(stat.)} \pm 17  ~\mathrm{(syst.)} \pm 6 ~ \mathrm{(lumi.)}\unit{pb},
\end{equation}

which is in good agreement with both the separate channel measurements and those from the
cross-check analyses discussed below. The corresponding summary of  the systematic uncertainties is given in Table~\ref{table:SHYFT_syssumm}. We obtain a result of $96~^{+5}_{-4}$\% for the $\cPqb$-tag scale factor which agrees well with the result obtained in \cite{CMS_btag}. For the jet energy scale we obtain a result of $107\pm 6$\% indicating that the data may prefer a small increase in the jet energy calibration. The scale factors for the $\PW$+$\cPqb$ jets and $\PW$+$\cPqc$ jets components indicate that the contributions to the data may be larger than what is predicted by the scaled NNLO predictions of 110\unit{pb} and 3.0\unit{nb}. For the $\PW$+$\cPqb$ jets contribution we find a cross section scale factor of 1.9 $^{+0.6}_{-0.5}$, which is similar to recent observations at the Tevatron \cite{new_d0, new_cdf, cdf_wb}. The result for the $\PW$+$\cPqc$ jets contribution is 1.4 $\pm$ 0.2.

\section{Cross-check Analyses}

As a cross-check of our results, we have performed a series of independent analyses in both the muon and electron channels. A summary of these is given in the following subsections. These use not only different analysis techniques but also different methods to suppress the backgrounds from $\PW$ + jets and QCD multijet events. Each analysis requires at least three jets with $\PT > 25\GeV$ and has no requirement on the amount of missing transverse energy. In addition, two different tagging algorithms are used for the two analyses in the muon channel. The Neural Network analysis uses a track counting algorithm which counts the number of tracks nonassociable to the primary vertex \cite{CMS_btag} and the second analysis uses the muons from semi-leptonic $\cPqb$ decays to tag $\cPqb$ jets.  Each analysis has significantly different systematic uncertainties from the analysis presented above and  thus provides a good test of the robustness of the measurements. While there are some differences in the selected event samples, we do not attempt to combine the results because of the
substantial overlap.

\subsection{\texorpdfstring{Neural Network Analysis with a Track-counting $\cPqb$-tagger}{Neural Network Analysis with a Track-counting b-tagger}}
\label{sec:NN_mu}

The first cross-check is performed in the muon channel. It makes use of a multi-layer perceptron neural network to distinguish $\ttbar$ signal events from the backgrounds after requiring a muon and three jets to pass the selection criteria. The network discriminant is built from the analysis of three input variables:  the pseudorapidity $|\eta^{\mu}|$ of the muon, the distance $\Delta R_{12}$ in $\eta$--$\phi$ space between the two highest-$\PT$ jets in the event, and a boolean variable indicating the presence of at least one $\cPqb$-tagged jet. The network was trained using simulated samples of signal and background events and the cross section was determined by fitting the sum of signal and background templates to the data.

The signal discriminant was generated from a simulated $\ttbar$ sample which was corrected to
match the jet energy resolution observed in the data. An additional flavor-dependent correction was applied to the simulated jets to account for differences in $\cPqb$-tag efficiencies between data and simulation. The template shapes for QCD, $\PW$+jets, and $\cPZ$+jets were produced directly from control regions in the data. These were chosen using variables that are only very loosely correlated to the network parameters so that they do not bias the operation of the network.

\begin{figure}[ht]
  \begin{center}
    \includegraphics[width=0.49\textwidth]{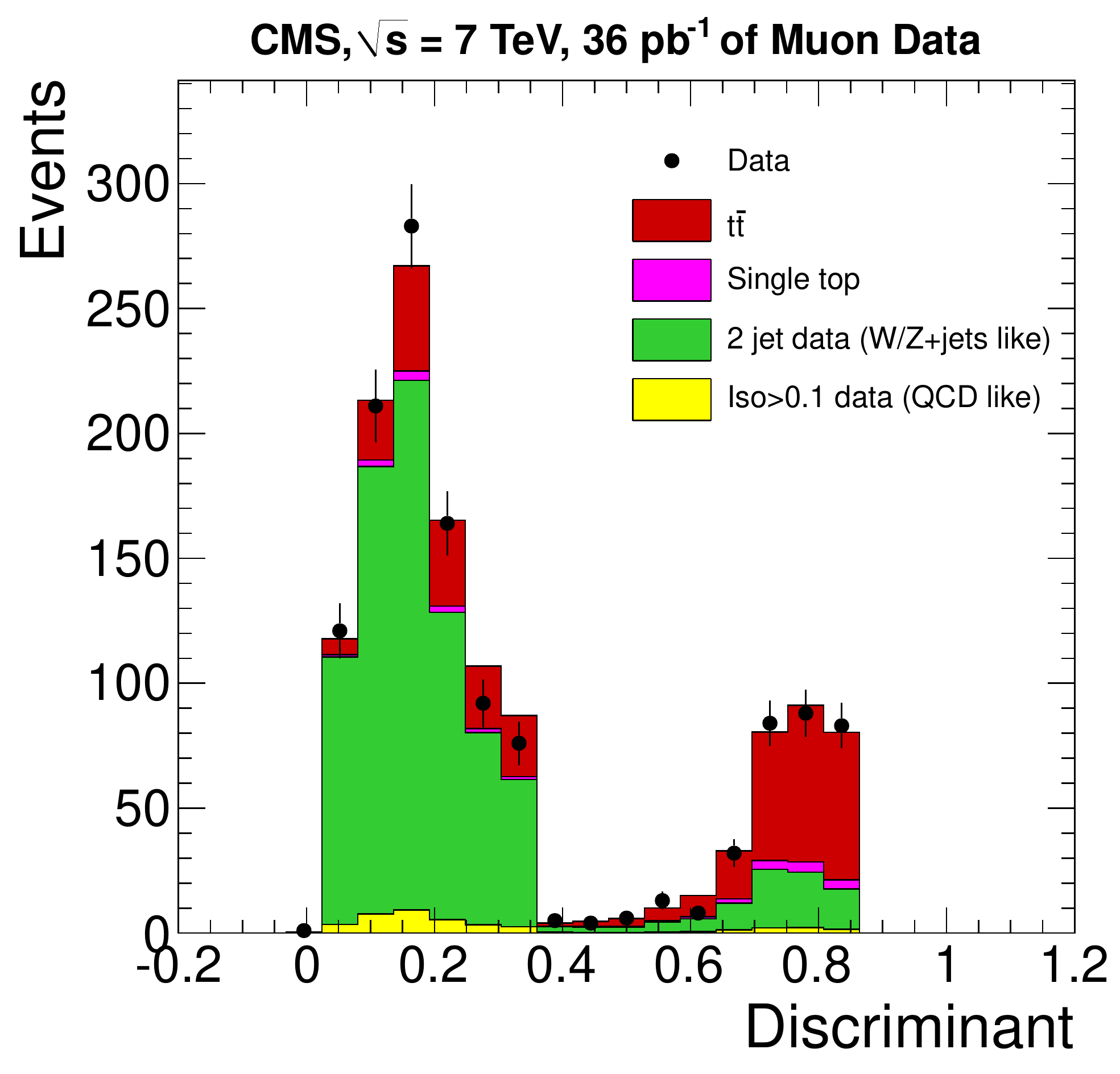}
    \caption{Results of the template fit to the neural network discriminant for 36\pbinv of data.}
    \label{fig:Steve_plot2}
  \end{center}
\end{figure}

Because the event topologies of $\ttbar$ and single-top events are so similar, there is very little difference in the shapes of the discriminant for the two samples. In order to avoid pathological fit yields
due to this similarity, the number of single-top events in the fit was constrained to the expected yield from the NLO single-top cross section with a 30\% uncertainty.  The QCD and $\PW/\cPZ$+jets likewise share similar shapes in their discriminants. This ambiguity was resolved by constraining the QCD fit yield according to its leading-order (LO) cross section, with a 100\% uncertainty due to expected differences between the actual QCD cross section and its LO calculation. However, as the QCD contribution is small, it has little effect on the final result.

The fit of the neural-network discriminant to events passing selection cuts in the data is shown in Fig.~\ref{fig:Steve_plot2}. A study using pseudo-experiments with simulated data indicates that the fitter introduces a $-3.1\%$ bias in the $\ttbar$ yield. This arises due to the use of control regions for the QCD, $\PW$+jets, and $\cPZ$+jets templates. After correcting for this, we obtain a $\ttbar$ event yield of 369~$\pm$~36~(stat.) events.

The systematic uncertainties are evaluated by shifting the simulation by each systematic uncertainty and re-evaluating the result. The resulting pseudo-data are fit with the nominal discriminator templates for $\ttbar$ and single top, and systematically shifted templates for QCD and $\PW/\cPZ$+jets. The dominant uncertainty of $^{+16}_{-15}\%$ comes from the $\cPqb$-tag efficiencies. When combined with the other contributions, we find a combined systematic uncertainty of $^{+23}_{-18}\%$. After combining this with the 4\% uncertainty in the recorded integrated luminosity~\cite{CMS_Lumi}, we obtain a $\ttbar$ cross section of 151~$\pm15$~(stat.)~$^{+35}_{-28}$~(syst.)~$\pm6$~(lumi.)\unit{pb}.

\subsection{\texorpdfstring{Muon Channel Analysis using a Muon-in-Jet $\cPqb$-tagger}{Muon Channel Analysis using a Muon-in-Jet b-tagger}}
\label{sec:SMT}

We have also performed an analysis in which $\cPqb$ jets are identified by the presence of a nonisolated muon. The event selection requires at least three well reconstructed jets of which at least
one contains a muon with  $\PT>$ 4\gev and $|\eta|<$ 2.4. The backgrounds from events which come from the decay of a $\JPsi$, $\Upsilon$, or $\cPZ$ boson, are excluded using selections on the muon pair invariant mass.

The background from $\PW$+$\cPqb$ jets and $\PW$+$\cPqc$ jets with a semimuonic decay and $\PW$+light flavor events with a jet that is misidentified as containing a semimuonic decay are estimated from data.
We calculate the track taggability from a sample of $\gamma$+jet events.  This
taggability is convoluted with the track distribution in the jets of the pre-tagged event sample to predict the number of tagged events that should arise from $\PW$+jets, $\mathrm{N}_{\text{tag,pred}}$.  We correct for the fraction of events in this
sample that are due to QCD multijets, $F_{\text{QCD}}$, to obtain
$N_{\PW+\mathrm{jets}} = N_{\text{tag,pred}}\cdot(1 - F_{\text{QCD}})$.
We estimate $F_{\mathrm{QCD}}$ using the method discussed below. We assign a 30\% systematic uncertainty, based on studies of the taggability parameterization as applied to independent control samples, to the tag-rate prediction to account for the simulation uncertainties.

The QCD background was estimated by calculating the fraction of QCD events in the pre-tag sample. Because of the enrichment in $\cPqb\cPaqb$ and $\cPqc\cPaqc$ events after requiring the primary muon, a correction factor $k$ is applied to correct for this. The QCD background is then  $N_{\rm{QCD}} = N_{\rm{tag,pred}} \cdot F_{\rm{\rm{QCD}}} \cdot k$. After taking into account the uncertainties in the calculation, we assign a systematic uncertainty in the QCD background of 60\%.

The background from Drell--Yan events that survive the $\cPZ$ veto is estimated from simulation.  An estimate from data similar to the one described in reference~\cite{Acosta:2005zd} was employed to assign a systematic uncertainty to the prediction from simulation.  This gives a systematic uncertainty of 17\% on the $\cPZ$ mass veto correction. The remaining backgrounds from diboson and single-top production were estimated from simulated samples that were normalized to the theoretical NLO cross sections. Each of these is assigned a 30\% systematic uncertainty.

The selection efficiency for signal events prior to the muon-tag requirement is evaluated using simulated $\ttbar$ events and the method described previously. The combined systematic uncertainties on the acceptance, due to the simulation uncertainties, is $^{+6.1}_{-6.9}$\%. Because of the relatively soft $\PT$ spectrum of the tag muons, the reconstruction and identification efficiencies were checked using a tag-and-probe analysis~\cite{CMS_WZ} of $\JPsi\rightarrow\Pgmp\Pgmm$ events. The results agree with the simulation to within 1\%, so we take the efficiency for finding a tag muon and the mistag efficiency directly from simulation. This gives a tagging efficiency for $\ttbar$ events of 25.4~$\pm$~0.1\% excluding the resonance veto requirements, and 23.9~$\pm$~0.1\% including them. The systematic uncertainty on this efficiency is conservatively taken to be 10\%. Before calculating the $\ttbar$ cross section, the number of predicted tagged events is corrected for the presence of $\ttbar$ events in the pre-tag sample. This correction is performed
iteratively. Combining these effects gives a total systematic uncertainty on the $\ttbar$ cross section due to the background calculations of 17.9\%.

\begin{figure}[hbt]
\begin{center}
\includegraphics[width=0.45\textwidth]{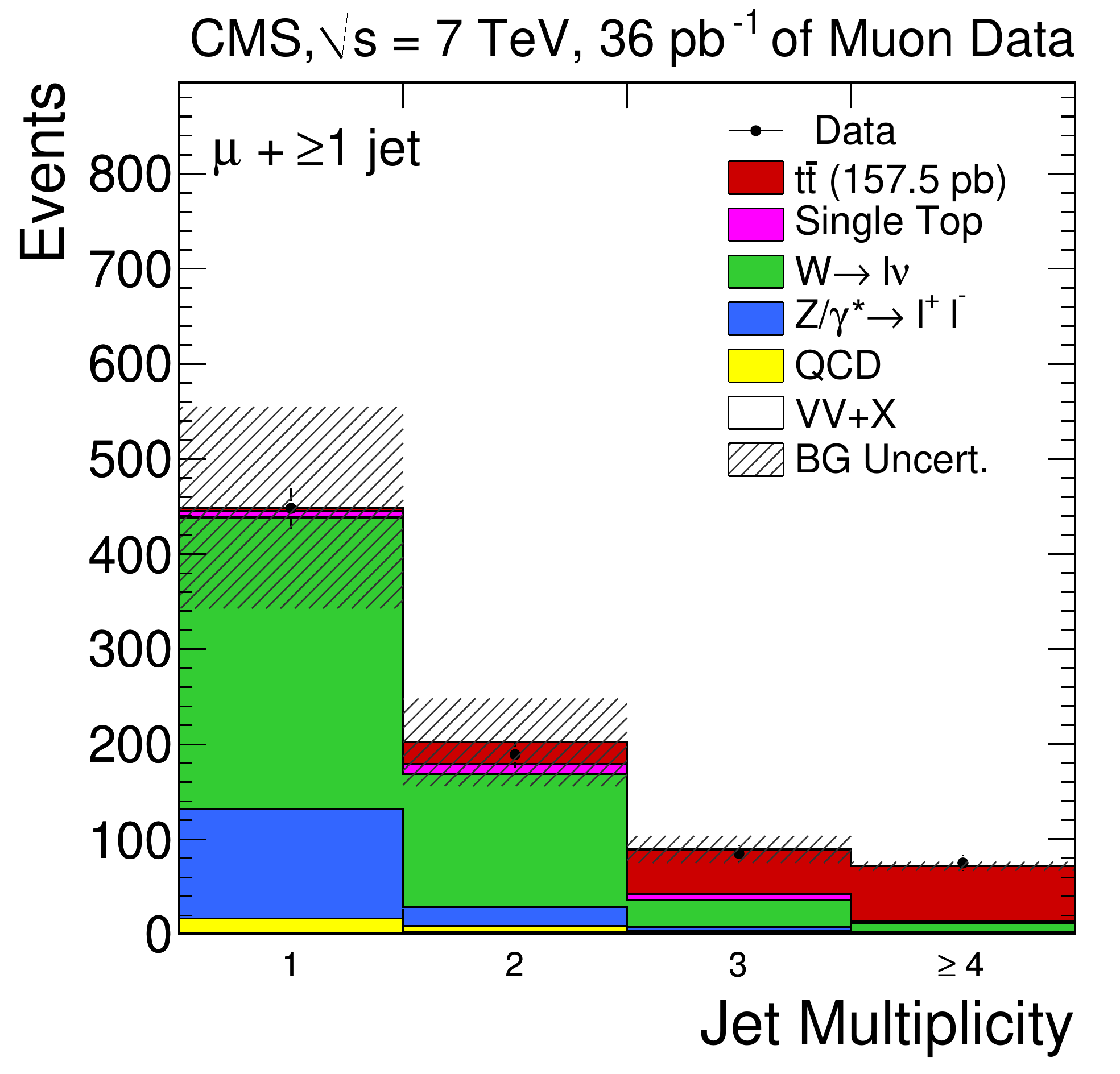}
\caption{Comparison of data and simulation for the jet multiplicity for events with $\ge$ 1 muon-in-jet tag. The $\PW$+jets and QCD contributions are normalized to the data-driven predictions. The hatched area shows total uncertainty on the background prediction.
\label{pasfig:njet}}
\end{center}
\end{figure}

Figure~\ref{pasfig:njet} shows the observed jet multiplicity distribution, together with the signal and background predictions. Here, the $\ttbar$ component is normalized to the NLO cross section~\cite{Top_x1} and the backgrounds are those computed above. The combined signal and background predictions are in good agreement with the data for each jet multiplicity.

Combining the data yield, the background estimation, the $\ttbar$ acceptance,  and the tagging efficiency with all of the associated uncertainties, gives a cross section result of 163 ~$\pm$~21 (stat.) ~$\pm$~35~(syst.) $\pm$~7 (lumi.)\unit{pb}. This measurement provides a valuable cross-check of the muon+jets result, as its systematic uncertainties are almost independent of those in the reference analysis.

\subsection {Electron Channel Cross-Check Analysis}
\label{Sec:BCBR_e}

Our third cross-check analysis is a simple counting analysis using the same displaced vertex tagger as our reference analysis. The event selection requires the presence of at least three well reconstructed jets, of which one or more is required to be $\cPqb$ tagged. We estimate the QCD background from the data using control regions in isolation and the $\PW$+jets background from a combination of data and simulation. To calculate the $\ttbar$ event yield, these are subtracted from the data, along with smaller contributions from single-top, Drell--Yan, and diboson production, which are estimated from simulation.

\begin{figure}[htbp]
  \begin{center}
    \includegraphics[width=0.45\textwidth]{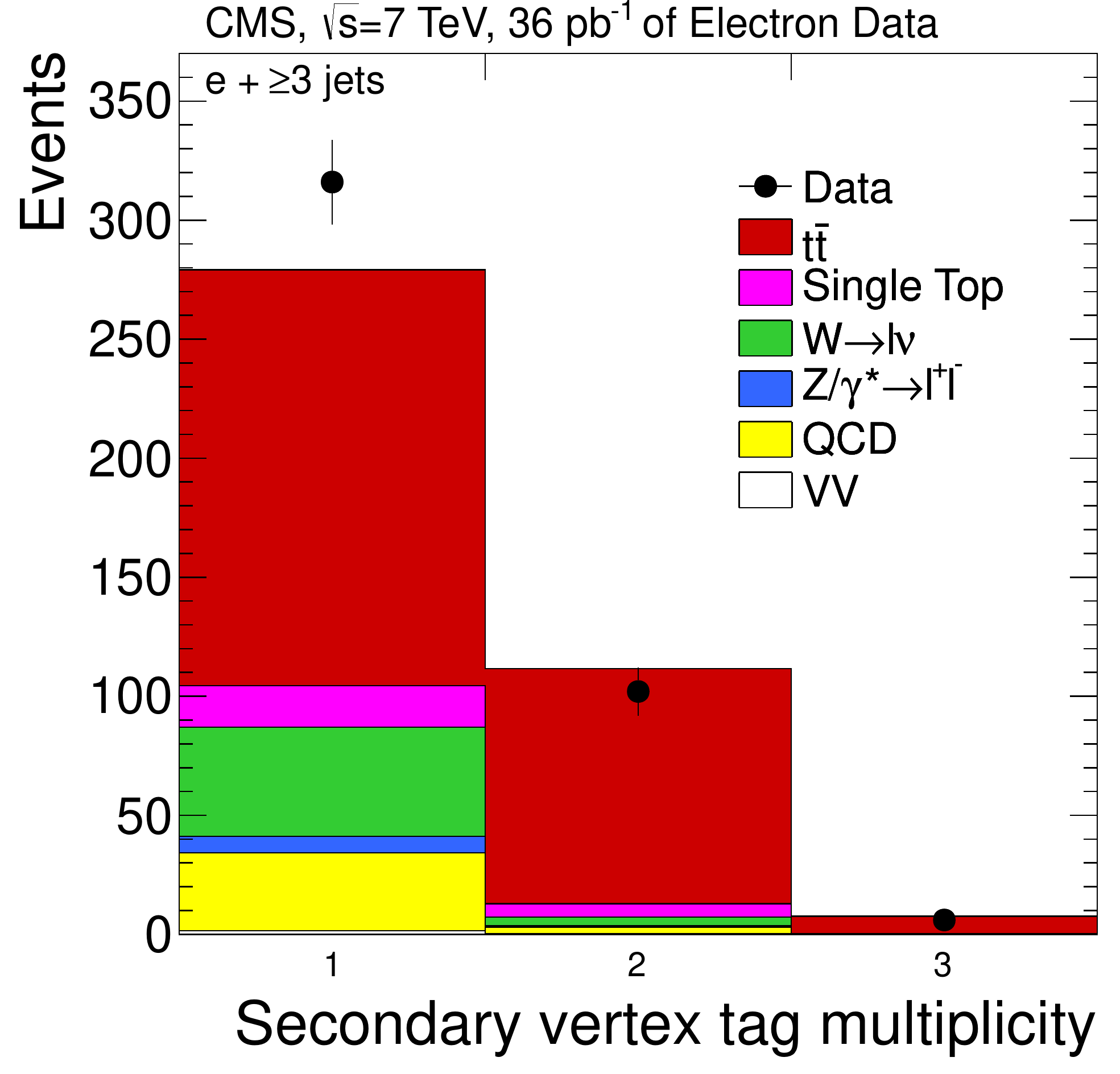}
    \caption{Number of observed and predicted secondary vertex tags in events with $\ge$ 3 jets.}
     \label{fig:tag_compare}
  \end{center}
\end{figure}

The selection efficiency is calculated for $\ttbar$ and for $\PW$+jets, and the efficiencies to tag individual jets are measured in bins of jet-$\PT$, separated into $\cPqb$, $\cPqc$, and $\cPqu,\cPqd,\cPqs,\cPg$ (light) partons. The uncertainties from the data are used in systematic studies as well as the statistical uncertainty from the jet efficiency calculations.  The resulting predicted $\cPqb$-tag rates applied to simulation of signal and backgrounds are in good agreement with the data (see Fig.~\ref{fig:tag_compare}).

The QCD background estimation is performed using a fit to the electron $I_{\mathrm{rel}}$ distribution in the data above the standard selection where the QCD multijet events dominate. A Gaussian function is used for the central value as it best fits the data, and the fit uncertainty is estimated by varying the $I_{\mathrm{rel}}$ range used in the fit and by using alternate fit functions. The systematic uncertainty from $\ttbar$ contamination is evaluated by subtracting the number of $\ttbar$ events predicted to be inside the fit region using simulation.

The background due to $\PW$+jets is estimated using a technique motivated by Berends--Giele scaling~\cite{Berends199132} to measure the jet multiplicity distribution before $\cPqb$-tagging.
In strict Berends--Giele scaling, the ratio $C(n)$ of events with $\PW + \geq n$ jets to events with $W +\geq (n+1)$ jets is expected to be independent of $n$.  To account for an observed deviation from perfect scaling, in this estimate, $C(n)$ is extracted as a linear function of $n$, with slope taken from the simulated event sample for $1 \leq n \leq 3$. The $\PW$+$\cPqb$ jets content in the simulation is scaled up by a factor of two, as determined in Section~\ref{sec:shyft_combined}.

A $\PW$+jets jet multiplicity spectrum in data is prepared by applying an additional requirement on the transverse missing energy, $\ETslash > 20$ GeV, to suppress the QCD background. The number of QCD events remaining in these data are then subtracted by refitting the $I_{\mathrm{rel}}$ distribution as above. The remaining  non-$\PW$ backgrounds and the $\ttbar$ content are subtracted from the data sample using Monte Carlo predictions. The normalization of $C(n)$ is then fitted to this data sample. The scale factors obtained are $C(1)=4.91\pm0.13$ and $C(2)=5.35\pm0.16$, where the uncertainties are statistical.  Using $C(n)$ the estimate of the $\PW$+jets content in the data with $\geq3$ jets before $\cPqb$ tagging ($N^{\mathrm{pretag, data}}$) is

\begin{eqnarray}
N^\mathrm{pretag, data}_{\PW+\mathrm{jets}, \ge3 \mathrm{jets}} = \frac{N^\mathrm{pretag, data}_{\PW+\mathrm{jets},\ge1 \mathrm{jets}}}{C(1)C(2)}.\label{eq:pre-taggedWJ_3jet}
\end{eqnarray}
The number of tagged $\PW$+jet events is then found by applying the selection efficiencies described above.

Systematic uncertainties for the $\PW$+jets background estimate are derived by repeating the above process varying $\ttbar$ content by 30\% as well as varying the normalization of $\PW$+$\cPqb$ jets by 50\% and $\PW$+$\cPqc$ jets content by factors of 2 and 0.5.  Additional fits are also done to Monte Carlo samples where the factorization scale is doubled or halved, as a systematic study.
The backgrounds from single-top, Drell-Yan, and diboson production are taken from simulation with a 30\% systematic uncertainty.

The background-subtracted and efficiency-corrected yield gives a cross section measurement of $169 \pm 13 (\rm{stat.})~^{+39}_{-32}~(\rm{syst.})~\pm 7~(\rm{lumi.})$\unit{pb}. The dominant systematic uncertainties are the uncertainties on the $\cPqb$-tag scale factors, and the uncertainty on the jet energy scale. This analysis uses a significantly different analysis strategy from the the reference electron+jets analysis but obtains a very similar cross section. Thus it provides a good cross-check of the method and the results.

\subsection{Additional Cross-Check}

As a further cross-check, the $\ttbar$ cross section was also measured in the muon channel using a simple counting analysis which used the same vertex tagger as our reference analysis. The analysis is similar to that of the electron channel analysis described in the previous subsection. The data were selected using a relaxed $I_{\mathrm{rel}}$ cut of $< 0.1$ and a missing transverse energy requirement of $\ETslash >$ 20\GeV. The QCD multijet background was measured from pre-tagged data using the matrix method \cite{D0MM} and the $\PW$ + jet background was derived from background-and-signal-subtracted data using the Berends--Giele method described in the Section~\ref{Sec:BCBR_e}. After requiring at least three well reconstructed jets with at least one jet with a $\cPqb$ tag and applying a correction of factor of 2 $\pm$ 1 for the $\PW$ + heavy flavor content of the simulated data, we find good agreement between the data and simulations. The resulting cross section is in good agreement with our reference result.

\section{Combined CMS Measurement}

In addition to the results from this analysis, CMS has also performed a measurement in the dilepton decay channel \cite{CMS_dilep}, where we measured a cross section of

\begin{center}
dileptons: $\sigma_{\ttbar} = 168  \pm 18  ~\mathrm{(stat.)} \pm 14  ~\mathrm{(syst.)} \pm 7  ~\mathrm{(lumi.)}$\unit{pb}.
\end{center}

To produce a final CMS result, we combine this result with the lepton + jets measurement using a profile likelihood method. This procedure uses the vertex mass templates information from this analysis and adds the dilepton measurements as single bin templates. For the dilepton channels, six statistically independent inputs are used so the uncertainty correlations can be handled correctly. These are the $\Pe\Pgm$ events with $\ge$ 2 jets and no $\cPqb$ tag requirement, the $\Pe\Pgm$, $\Pe\Pe$, and $\Pgm\Pgm$ events with only 1 jet, and the $\Pe\Pe$ and $\Pgm\Pgm$ events with $\ge$ 2 jets and at least 1 $\cPqb$ tag. Because different $\cPqb$ taggers were used in the two analyses, the $\cPqb$-tag uncertainties are treated as uncorrelated. The 3\% PDF uncertainty from the lepton + jets measurement is also treated as uncorrelated because it does not relate to anything in the dilepton analysis. The remaining systematic uncertainties (JES, lepton efficiency, renormalization and factorization scales, ME to PS matching, and ISR/FSR) are all assumed to be correlated. The method was verified by using pseudo-experiments, from which it was determined that the likelihood calculation resulted in a 10\% underestimate of the errors. We correct for the underestimate in the final result and obtain a combined measurement of

\begin{center}
CMS combined: $\sigma_{\ttbar} = 154  \pm 17  ~\mathrm{(stat.+syst.)} \pm 6  ~\mathrm{(lumi.)}$\unit{pb}.
\end{center}

\begin{figure}[htbp]
  \begin{center}
    \includegraphics[width=0.48\textwidth]{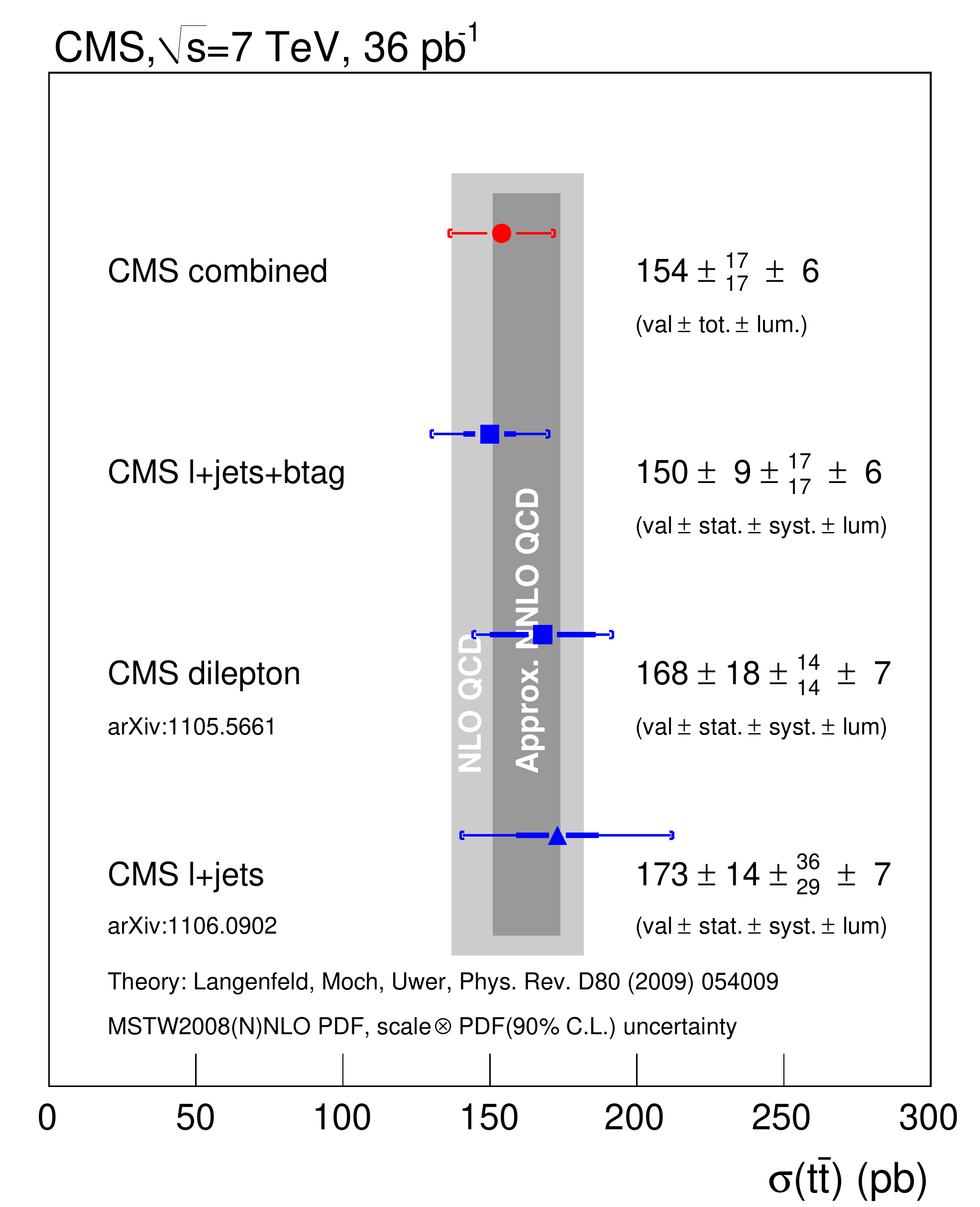}
    \caption[Expt Comparisions]{Comparison of the CMS measurements for the $\ttbar$ production cross  sections and the QCD predictions for $\sqrt{s}=7\TeV$}
    \label{fig:Result_compare1}
  \end{center}
\end{figure}

Figure~\ref{fig:Result_compare1} shows the comparison of the separate and combined CMS \cite{CMS_dilep,lepjets} measurements of the production cross section. The inner error bars on the data points correspond to the statistical uncertainty, while the outer (thinner) error bars correspond to the quadratic sum of the statistical and systematic uncertainties. The outermost brackets correspond to the total uncertainty, including a luminosity uncertainty of  11\% (4\%) for the 3 (36)\pbinv results, respectively, which is also added in quadrature.

 Also shown are NLO and approximate NNLO QCD calculations, for comparison. These were computed using the HATHOR program \cite{Aliev:2010zk} using the calculations from \cite{Langenfeld:2009wd}. A common factorization and renormalization scale of $Q = m_{\cPqt} = 173\GeV$ was used for the calculations together with the MRTSW 2008 NNLO (NLO) parton distribution functions. The scale uncertainty was determined by independently varying the two scales by factors of 2 and 0.5 and taking the maximum variation as the uncertainty. The PDF uncertainty corresponds to the 90\% confidence level (C.L.) uncertainties for the parton distribution functions \cite{pdf4lhc}. This is added in quadrature to the scale uncertainty.

\begin{figure}[ht]
  \begin{center}
    \includegraphics[angle=90,width=0.5\textwidth]{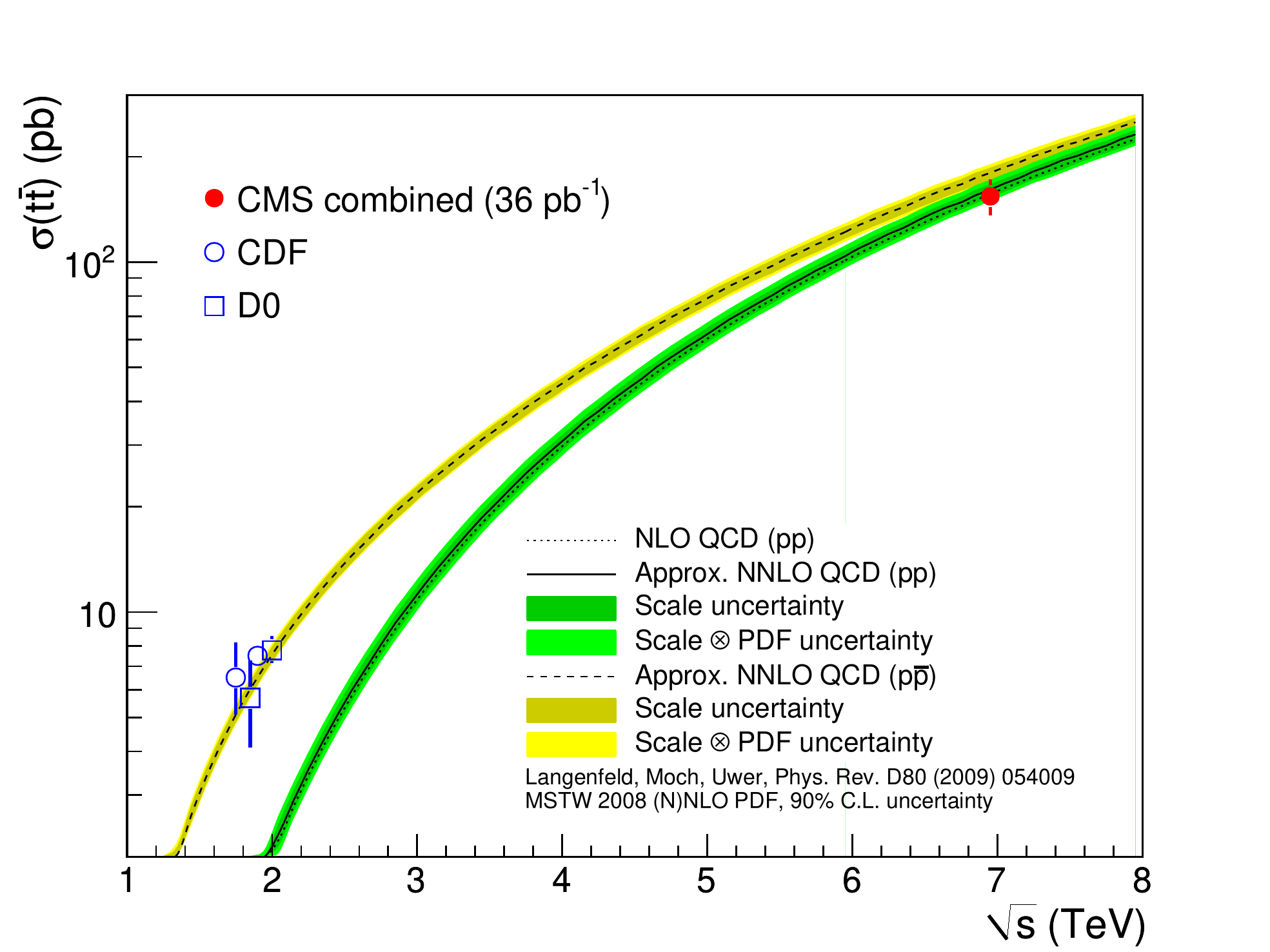}
    \caption[Thy Comparisions]{Comparison of the CMS and Tevatron results for the $\ttbar$ production cross section with QCD as a function of $\sqrt{s}$}
    \label{fig:Result_compare2}
  \end{center}
\end{figure}

In Fig.~\ref{fig:Result_compare2} we compare these and the $\Pp\Pap$ results from the Tevatron  \cite{old_d0,old_cdf,*old_cdf_erratum,new_d0,new_cdf} to the theoretical predictions as a function of $\sqrt{s}$. These are from the approximate NNLO QCD calculations referred to above and the width of the error band corresponds to the variation resulting from changing the $Q^{2}$ scale up and down by a factor of 2.  We find good agreement between the data and the theory in all cases and we note that the combined measurement is already more precise than the NLO QCD prediction.

\section{Summary of Results}

We have presented the results of a new analysis of the $\ttbar$ production cross section at $\sqrt{s}=7\,\rm{TeV}$ using data recorded by the CMS detector at the LHC during 2010 corresponding to an integrated luminosity of 36\pbinv. Using muon and electron+jets and using $\cPqb$ tagging to suppress the backgrounds, we measure cross sections of

\begin{center}
$\Pgm$+jets: $\sigma_{\ttbar} = 145  \pm 12  ~\mathrm{(stat.)} \pm 18  ~\mathrm{(syst.)} \pm 6  ~\mathrm{(lumi.)}$\unit{pb},

$\Pe$+jets:  $\sigma_{\ttbar} = 158  \pm 14  ~\mathrm{(stat.)} \pm 19  ~\mathrm{(syst.)} \pm 6  ~\mathrm{(lumi.)}$\unit{pb},
\end{center}

from the separate channels. The combination of these gives a cross section of

\begin{center}
$l$+jets: $\sigma_{\ttbar} = 150  \pm 9  ~\mathrm{(stat.)} \pm 17  ~\mathrm{(syst.)} \pm 6  ~\mathrm{(lumi.)}$\unit{pb}.
\end{center}

When combined with the CMS dilepton measurement, we obtain an improved cross section measurement of

\begin{center}
CMS combined: $\sigma_{\ttbar} = 154  \pm 17  ~\mathrm{(stat.+syst.)} \pm 6  ~\mathrm{(lumi.)}$\unit{pb}.
\end{center}

The measurements are in good agreement with the QCD predictions of 164$^{+10}_{-13}$\unit{pb} \cite{Aliev:2010zk,Langenfeld:2009wd} and 163$^{+11}_{-10}$\unit{pb} \cite{Kidonakis:2010dk} which are based on the full NLO matrix elements and the resummation of the leading and next-to-leading soft logarithms.

\section*{Acknowledgements}
We wish to congratulate our colleagues in the CERN accelerator departments for the excellent performance of the LHC machine. We thank the technical and administrative staff at CERN and other CMS Institutes, and acknowledge support from: FMSR (Austria); FNRS and FWO (Belgium); CNPq, CAPES, FAPERJ, and FAPESP (Brazil); MES (Bulgaria); CERN; CAS, MoST, and NSFC (China); COLCIENCIAS (Colombia); MSES (Croatia); RPF (Cyprus); Academy of Sciences and NICPB (Estonia); Academy of Finland, ME, and HIP (Finland); CEA and CNRS/IN2P3 (France); BMBF, DFG, and HGF (Germany); GSRT (Greece); OTKA and NKTH (Hungary); DAE and DST (India); IPM (Iran); SFI (Ireland); INFN (Italy); NRF (Korea); LAS (Lithuania); CINVESTAV, CONACYT, SEP, and UASLP-FAI (Mexico); PAEC (Pakistan); SCSR (Poland); FCT (Portugal); JINR (Armenia, Belarus, Georgia, Ukraine, Uzbekistan); MST and MAE (Russia); MSTDS (Serbia); MICINN and CPAN (Spain); Swiss Funding Agencies (Switzerland); NSC (Taipei); TUBITAK and TAEK (Turkey); STFC (United Kingdom); DOE and NSF (USA). Individuals have received support from the Marie-Curie IEF program (European Union); the Leventis Foundation; the A. P. Sloan Foundation; the Alexander von Humboldt Foundation; and the World Class University program by NRF (Korea).

\bibliography{auto_generated}

\providecommand{\href}[2]{#2}\begingroup\raggedright\begin{thebibliography}{10}%
\makeatletter
\providecommand{\hrefCMSnoop }[0]{\@secondoftwo}%
\makeatother

\bibitem{top-discovery-cdf}
\hrefCMSnoop {} {{ CDF} Collaboration, ``{Observation of top quark production
  in ${\rm \bar{p}p}$ collisions}'',} \textit{ Phys. Rev. Lett.} \textbf{ 74}
  (1995) 2626, \href{http://www.arXiv.org/abs/9503002}{\texttt{
  arXiv:9503002}}.
\href{http://dx.doi.org/10.1103/PhysRevLett.74.2626}{\texttt{
  doi:10.1103/PhysRevLett.74.2626}}.

\bibitem{top-discovery-d0}
\hrefCMSnoop {} {{ D0} Collaboration, ``{Observation of the top quark}'',}
  \textit{ Phys. Rev. Lett.} \textbf{ 74} (1995) 2632,
  \href{http://www.arXiv.org/abs/9503003}{\texttt{ arXiv:9503003}}.
\href{http://dx.doi.org/10.1103/PhysRevLett.74.2632}{\texttt{
  doi:10.1103/PhysRevLett.74.2632}}.

\bibitem{top-prospects}
\hrefCMSnoop {} {J.~R. Incandela {et~al.}, ``{Status and Prospects of Top-Quark
  Physics}'',} \textit{ Prog. Part. Nucl. Phys.} \textbf{ 63} (2009) 239,
  \href{http://www.arXiv.org/abs/0904.2499}{\texttt{ arXiv:0904.2499}}.
\href{http://dx.doi.org/10.1016/j.ppnp.2009.08.001}{\texttt{
  doi:10.1016/j.ppnp.2009.08.001}}.

\bibitem{lhc}
\hrefCMSnoop {} {L.~Evans and P.~Bryant~(editors), ``{LHC Machine}'',} \textit{
  JINST} \textbf{ 03} (2008) S08001.
\href{http://dx.doi.org/10.1088/1748-0221/3/08/S08001}{\texttt{
  doi:10.1088/1748-0221/3/08/S08001}}.

\bibitem{Aliev:2010zk}
\hrefCMSnoop {} {M.~Aliev {et~al.}, ``{HATHOR: HAdronic Top and Heavy quarks
  crOss section calculatoR}'',} \textit{ Comput. Phys. Commun.} \textbf{ 182}
  (2011) 1034, \href{http://www.arXiv.org/abs/1007.1327}{\texttt{
  arXiv:1007.1327}}.
  \href{http://dx.doi.org/10.1016/j.cpc.2010.12.040}{\texttt{
  doi:10.1016/j.cpc.2010.12.040}}.

\bibitem{Langenfeld:2009wd}
\hrefCMSnoop {} {U.~Langenfeld, S.~Moch, and P.~Uwer, ``{Measuring the running
  top-quark mass}'',} \textit{ Phys. Rev. D} \textbf{ 80} (2009) 054009,
  \href{http://www.arXiv.org/abs/0906.5273}{\texttt{ arXiv:0906.5273}}.
  \href{http://dx.doi.org/10.1103/PhysRevD.80.054009}{\texttt{
  doi:10.1103/PhysRevD.80.054009}}.

\bibitem{Kidonakis:2010dk}
\hrefCMSnoop {} {N.~Kidonakis, ``{Next-to-next-to-leading soft-gluon
  corrections for the top quark cross section and transverse momentum
  distribution}'',} \textit{ Phys. Rev. D} \textbf{ 82} (2010) 114030,
  \href{http://www.arXiv.org/abs/1009.4935}{\texttt{ arXiv:1009.4935}}.
  \href{http://dx.doi.org/10.1103/PhysRevD.82.114030}{\texttt{
  doi:10.1103/PhysRevD.82.114030}}.

\bibitem{CMS_Lumi}
\href {http://cdsweb.cern.ch/record/1279145} {{ CMS} Collaboration,
  ``Measurement of {CMS} Luminosity'',} CMS Physics Analysis Summary
  CMS-PAS-EWK-10-004, (2010).

\bibitem{dilep}
\hrefCMSnoop {} {{ CMS} Collaboration, ``First Measurement of the Cross Section
  for Top-Quark Pair Production in Proton-Proton Collisions at 7 TeV'',}
  \textit{ Phys. Lett. B} \textbf{ 695} (2011) 424.
\href{http://dx.doi.org/10.1016/j.physletb.2010.11.058}{\texttt{
  doi:10.1016/j.physletb.2010.11.058}}.

\bibitem{lepjets}
\hrefCMSnoop {} {{ CMS} Collaboration, ``Measurement of the \ttbar Pair
  Production Cross Section using the Kinematic Properties of Lepton + Jets
  Events'',} (2011). \href{http://www.arXiv.org/abs/1106.0902}{\texttt{
  arXiv:1106.0902}}. Submitted to \textit{Eur. Phys. J. C}.

\bibitem{ATLAS:2010}
\hrefCMSnoop {} {{ ATLAS} Collaboration, ``Measurement of the top quark-pair
  production cross section with {ATLAS} in pp collisions at $\sqrt{s}$ = 7
  {TeV}'',} (2010). \href{http://www.arXiv.org/abs/1012.1792}{\texttt{
  arXiv:1012.1792}}. Accepted by \textit{Eur. Phys. J. C}.

\bibitem{cms}
\hrefCMSnoop {} {{ CMS} Collaboration, ``The {CMS} experiment at the {CERN}
  {LHC}'',} \textit{ JINST} \textbf{ 3} (2008) S08004.
\href{http://dx.doi.org/10.1088/1748-0221/3/08/S08004}{\texttt{
  doi:10.1088/1748-0221/3/08/S08004}}.

\bibitem{CMS_mu}
\href {http://cdsweb.cern.ch/record/1279140} {{ CMS} Collaboration,
  ``Performance of muon identification in pp collisions at $\sqrt{s}$ = 7
  {TeV}'',} CMS Physics Analysis Summary CMS-PAS-MUO-10-002, (2010).

\bibitem{CMS_e}
\href {http://cdsweb.cern.ch/record/1299116} {{ CMS} Collaboration, ``Electron
  Reconstruction and Identification at $\sqrt{s} = 7$ {TeV}'',} CMS Physics
  Analysis Summary CMS-PAS-EGM-10-004, (2010).

\bibitem{CMS_vertex}
\hrefCMSnoop {} {{ CMS} Collaboration, ``CMS Tracking Performance Results from
  Early LHC Operation'',} \textit{ Eur. Phys. J. C} \textbf{ 70} (2010) 1165.
\href{http://dx.doi.org/10.1140/epjc/s10052-010-1491-3}{\texttt{
  doi:10.1140/epjc/s10052-010-1491-3}}.

\bibitem{ipartf}
\href {http://cdsweb.cern.ch/record/1279341} {{ CMS} Collaboration,
  ``Commissioning of the Particle-Flow Reconstruction in Minimum-Bias and Jet
  Events from {\Pp\Pp} Collisions at 7 {TeV}'',} CMS Physics Analysis Summary
  CMS-PAS-PFT-10-002, (2010).

\bibitem{ktalg}
\hrefCMSnoop {} {M.~Cacciari, G.~P. Salam, and G.~Soyez, ``{The anti-kt jet
  clustering algorithm}'',} \textit{ JHEP} \textbf{ 04} (2008) 063.
\href{http://dx.doi.org/10.1088/1126-6708/2008/04/063}{\texttt{
  doi:10.1088/1126-6708/2008/04/063}}.

\bibitem{fastjet1}
\hrefCMSnoop {} {M.~Cacciari and G.~P. Salam, ``{Dispelling the N**3 myth for
  the k(t) jet-finder}'',} \textit{ Phys. Lett. B} \textbf{ 641} (2006) 57,
  \href{http://www.arXiv.org/abs/hep-ph/0512210}{\texttt{
  arXiv:hep-ph/0512210}}.
\href{http://dx.doi.org/10.1016/j.physletb.2006.08.037}{\texttt{
  doi:10.1016/j.physletb.2006.08.037}}.

\bibitem{fastjet2}
\href {http://fastjet.fr} {M.~Cacciari, G.~P. Salam, and G.~Soyez, ``Fastjet
  Package'',} (2011).

\bibitem{jec}
\href {http://cdsweb.cern.ch/record/1369486} {{ CMS} Collaboration,
  ``Determination of the Jet Energy Scale in {CMS} with pp Collisions at
  $\sqrt{s}= 7$ {TeV}'',} CMS Physics Analysis Summary CMS-PAS-JME-10-010,
  (2010).

\bibitem{CMS_btag}
\href {http://cdsweb.cern.ch/record/1194494} {{ CMS} Collaboration,
  ``Algorithms for b Jet Identification in {CMS}'',} CMS Physics Analysis
  Summary CMS-PAS-BTV-09-001, (2009).

\bibitem{Alwall:2007st}
\hrefCMSnoop {} {J.~Alwall {et~al.}, ``{MadGraph/MadEvent v4: The New Web
  Generation}'',} \textit{ JHEP} \textbf{ 09} (2007) 028.
\href{http://dx.doi.org/10.1088/1126-6708/2007/09/028}{\texttt{
  doi:10.1088/1126-6708/2007/09/028}}.

\bibitem{Sjostrand:2006za}
\hrefCMSnoop {} {T.~Sj{\"o}strand, S.~Mrenna, and P.~Z. Skands, ``{PYTHIA 6.4
  Physics and Manual}'',} \textit{ JHEP} \textbf{ 05} (2006) 026,
  \href{http://www.arXiv.org/abs/0603175}{\texttt{ arXiv:0603175}}.
\href{http://dx.doi.org/10.1088/1126-6708/2006/05/026}{\texttt{
  doi:10.1088/1126-6708/2006/05/026}}.

\bibitem{Geant4}
\hrefCMSnoop {} {J.~Allison {et~al.}, ``{Geant4} developments and
  applications'',} \textit{ IEEE Trans. Nucl. Sci.} \textbf{ 53} (2006) 270.
\href{http://dx.doi.org/10.1109/TNS.2006.869826}{\texttt{
  doi:10.1109/TNS.2006.869826}}.

\bibitem{Top_x1}
\hrefCMSnoop {} {J.~Campbell and R.~Ellis, ``{MCFM} for the {T}evatron and the
  {LHC}'',} (2010). \href{http://www.arXiv.org/abs/1007.3492}{\texttt{
  arXiv:1007.3492}}.


\bibitem{mstw08}
\hrefCMSnoop {} {A.~D. Martin, W.~J. Stirling, R.~S. Thorne{ et~al.},
  ``{Uncertainties on $\alpha_s$ in global PDF analyses and implications for
  predicted hadronic cross sections}'',} \textit{ Eur. Phys. J. C} \textbf{ 64}
  (2009) 653, \href{http://www.arXiv.org/abs/0905.3531}{\texttt{
  arXiv:0905.3531}}.
\href{http://dx.doi.org/10.1140/epjc/s10052-009-1164-2}{\texttt{
  doi:10.1140/epjc/s10052-009-1164-2}}.

\bibitem{cteq_2010}
\hrefCMSnoop {} {H.-L. Lai {et~al.}, ``{Uncertainty induced by QCD coupling in
  the CTEQ-TEA global analysis of parton distributions}'',} \textit{ Phys. Rev.
  D} \textbf{ 82} (2010) 054021,
  \href{http://www.arXiv.org/abs/1004.4624}{\texttt{ arXiv:1004.4624}}.
  \href{http://dx.doi.org/10.1103/PhysRevD.82.054021}{\texttt{
  doi:10.1103/PhysRevD.82.054021}}.

\bibitem{nnpdf}
\hrefCMSnoop {} {F.~Demartin, S.~Forte, E.~Mariani{ et~al.}, ``{The impact of
  PDF and $\alpha_s$ uncertainties on Higgs Production in gluon fusion at
  hadron colliders}'',} \textit{ Phys. Rev. D} \textbf{ 82} (2010) 014002,
  \href{http://www.arXiv.org/abs/1004.0962}{\texttt{ arXiv:1004.0962}}.
\href{http://dx.doi.org/10.1103/PhysRevD.82.014002}{\texttt{
  doi:10.1103/PhysRevD.82.014002}}.

\bibitem{pdf4lhc}
\href {http://www.hep.ucl.ac.uk/pdf4lhc/PDF4LHCrecom.pdf} {{PDF4LHC Working
  Group}, ``{PDF4LHC} Recommendations'',} (2010).

\bibitem{mcfm2}
\hrefCMSnoop {} {J.~M. Campbell, R.~Frederix, F.~Maltoni{ et~al.},
  ``{Next-to-Leading-Order Predictions for t-Channel Single-Top Production at
  Hadron Colliders}'',} \textit{ Phys. Rev. Lett.} \textbf{ 102} (2009) 182003,
  \href{http://www.arXiv.org/abs/0903.0005}{\texttt{ arXiv:0903.0005}}.
\href{http://dx.doi.org/10.1103/PhysRevLett.102.182003}{\texttt{
  doi:10.1103/PhysRevLett.102.182003}}.

\bibitem{mcfm3}
\hrefCMSnoop {} {J.~M. Campbell and F.~Tramontano, ``{Next-to-leading order
  corrections to W t production and decay}'',} \textit{ Nucl. Phys. B} \textbf{
  726} (2005) 109, \href{http://www.arXiv.org/abs/0506289}{\texttt{
  arXiv:0506289}}.
\href{http://dx.doi.org/10.1016/j.nuclphysb.2005.08.015}{\texttt{
  doi:10.1016/j.nuclphysb.2005.08.015}}.

\bibitem{mcfm4}
\hrefCMSnoop {} {J.~M. Campbell, R.~K. Ellis, and F.~Tramontano, ``{Single top
  production and decay at next-to-leading order}'',} \textit{ Phys. Rev. D}
  \textbf{ 70} (2004) 094012,
  \href{http://www.arXiv.org/abs/hep-ph/0408158}{\texttt{
  arXiv:hep-ph/0408158}}.
\href{http://dx.doi.org/10.1103/PhysRevD.70.094012}{\texttt{
  doi:10.1103/PhysRevD.70.094012}}.

\bibitem{CMS_stop}
\href {http://cdsweb.cern.ch/record/1335719} {{ CMS} Collaboration,
  ``Measurement of the single-top $t$-channel cross section in pp collisions at
  $\sqrt{s}=7$~{TeV}'',} CMS Physics Analysis Summary CMS-PAS-TOP-10-008,
  (2010).

\bibitem{fewz}
\hrefCMSnoop {} {K.~Melnikov and F.~Petriello, ``Electroweak gauge boson
  production at hadron colliders through $O(\alpha_s^2)$'',} \textit{ Phys.
  Rev. D} \textbf{ 74} (2006) 114017,
  \href{http://www.arXiv.org/abs/hep-ph/0609070}{\texttt{
  arXiv:hep-ph/0609070}}.
\href{http://dx.doi.org/10.1103/PhysRevD.74.114017}{\texttt{
  doi:10.1103/PhysRevD.74.114017}}.

\bibitem{CMS_WZ}
\hrefCMSnoop {} {{ CMS} Collaboration, ``Measurements of Inclusive W and Z
  Cross Sections in pp Collisions at $\sqrt{s}$ = 7 TeV'',} \textit{ JHEP}
  \textbf{ 01} (2011) 080.
\href{http://dx.doi.org/10.1007/JHEP01(2011)080}{\texttt{
  doi:10.1007/JHEP01(2011)080}}.

\bibitem{CMS_btag2}
\href {http://cdsweb.cern.ch/record/1279144} {{ CMS} Collaboration,
  ``Commissioning of b-jet identification with pp collisions at $\sqrt{s}$ = 7
  {TeV}'',} CMS Physics Analysis Summary CMS-PAS-BTV-10-001, (2010).

\bibitem{new_d0}
\hrefCMSnoop {} {{ D0} Collaboration, ``{Measurement of the top quark pair
  production cross section in the lepton+jets channel in proton-antiproton
  collisions at $\sqrt{s}$ = 1.96 TeV}'',} \textit{ Phys. Rev. D} \textbf{ 84}
  (2011) 012008, \href{http://www.arXiv.org/abs/1101.0124}{\texttt{
  arXiv:1101.0124}}.
\href{http://dx.doi.org/10.1103/PhysRevD.84.012008}{\texttt{
  doi:10.1103/PhysRevD.84.012008}}.

\bibitem{new_cdf}
\href
  {http://www-cdf.fnal.gov/physics/new/top/confNotes/cdf9913_ttbarxs4invfb.ps}
  {{ CDF} Collaboration, ``Combination of CDF top quark pair production
  measurements with up to $4.6 \rm\ fb^{-1}$'',} CDF Note 9913, (2009).

\bibitem{cdf_wb}
\hrefCMSnoop {} {{ CDF} Collaboration, ``{First measurement of the $b$-jet
  cross section in events with a $W$ boson in proton-antiproton collisions at
  $\sqrt{s}$ = 1.96 TeV}'',} \textit{ Phys. Rev. Lett.} \textbf{ 104} (2010)
  131801, \href{http://www.arXiv.org/abs/0909.1505}{\texttt{ arXiv:0909.1505}}.
\href{http://dx.doi.org/10.1103/PhysRevLett.104.131801}{\texttt{
  doi:10.1103/PhysRevLett.104.131801}}.

\bibitem{Acosta:2005zd}
\hrefCMSnoop {} {{ CDF} Collaboration, ``{Measurement of the $t\bar{t}$
  production cross-section in $p\bar{p}$ collisions at $\sqrt{s} = 1.96\TeV$
  using lepton plus jets events with semileptonic B decays to muons}'',}
  \textit{ Phys. Rev. D} \textbf{ 72} (2005) 032002,
  \href{http://www.arXiv.org/abs/0506001}{\texttt{ arXiv:0506001}}.
  \href{http://dx.doi.org/10.1103/PhysRevD.72.032002}{\texttt{
  doi:10.1103/PhysRevD.72.032002}}.

\bibitem{Berends199132}
\hrefCMSnoop {} {F.~A. Berends {et~al.}, ``{On the production of a W and jets
  at hadron colliders}'',} \textit{ Nucl. Phys. B} \textbf{ 357} (1991) 32.
\href{http://dx.doi.org/10.1016/0550-3213(91)90458-A}{\texttt{
  doi:10.1016/0550-3213(91)90458-A}}.

\bibitem{D0MM}
\hrefCMSnoop {} {{ D0} Collaboration, ``{Measurement of the t anti-t production
  cross section in p anti-p collisions at $\sqrt{s} = 1.96\TeV$ using secondary
  vertex b tagging}'',} \textit{ Phys. Rev. D} \textbf{ 74} (2006) 112004,
  \href{http://www.arXiv.org/abs/0611002}{\texttt{ arXiv:0611002}}.
\href{http://dx.doi.org/10.1103/PhysRevD.74.112004}{\texttt{
  doi:10.1103/PhysRevD.74.112004}}.

\bibitem{CMS_dilep}
\hrefCMSnoop {} {{ CMS} Collaboration, ``Measurement of the top-quark pair
  production cross section and the top-quark mass in the dilepton channel at
  $\sqrt{s}$ = 7 TeV'',} \textit{ JHEP} \textbf{ 7} (2011) 49.
  \href{http://dx.doi.org/10.1007/JHEP07(2011)049}{\texttt{
  doi:10.1007/JHEP07(2011)049}}.

\bibitem{old_d0}
\hrefCMSnoop {} {{ D0} Collaboration, ``{t anti-t production cross-section in p
  anti-p collisions at $\sqrt{s} = 1.8\TeV$}'',} \textit{ Phys. Rev. D}
  \textbf{ 67} (2003) 012004, \href{http://www.arXiv.org/abs/0205019}{\texttt{
  arXiv:0205019}}. \href{http://dx.doi.org/10.1103/PhysRevD.67.012004}{\texttt{
  doi:10.1103/PhysRevD.67.012004}}.

\bibitem{old_cdf}
\hrefCMSnoop {} {{ CDF} Collaboration, ``{Measurement of the t anti-t
  production cross-section in p anti-p collisions at $\sqrt{s} = 1.8\TeV$}'',}
  \textit{ Phys. Rev. D} \textbf{ 64} (2001) 032002,
  \href{http://www.arXiv.org/abs/0101036}{\texttt{ arXiv:0101036}}.
  \href{http://dx.doi.org/10.1103/PhysRevD.64.032002}{\texttt{
  doi:10.1103/PhysRevD.64.032002}}.

\bibitem{old_cdf_erratum}
\hrefCMSnoop {} {{ CDF} Collaboration, ``Erratum: Measurement of the t anti-t
  production cross-section in p anti-p collisions at $\sqrt{s} = 1.8\TeV$'',}
  \textit{ Phys. Rev. D} \textbf{ 67} (2003) 119901(E).
  \href{http://dx.doi.org/10.1103/PhysRevD.67.119901}{\texttt{
  doi:10.1103/PhysRevD.67.119901}}.

\end{thebibliography}\endgroup

\cleardoublepage \appendix\section{The CMS Collaboration \label{app:collab}}\begin{sloppypar}\hyphenpenalty=5000\widowpenalty=500\clubpenalty=5000\textbf{Yerevan Physics Institute,  Yerevan,  Armenia}\\*[0pt]
S.~Chatrchyan, V.~Khachatryan, A.M.~Sirunyan, A.~Tumasyan
\vskip\cmsinstskip
\textbf{Institut f\"{u}r Hochenergiephysik der OeAW,  Wien,  Austria}\\*[0pt]
W.~Adam, T.~Bergauer, M.~Dragicevic, J.~Er\"{o}, C.~Fabjan, M.~Friedl, R.~Fr\"{u}hwirth, V.M.~Ghete, J.~Hammer\cmsAuthorMark{1}, S.~H\"{a}nsel, M.~Hoch, N.~H\"{o}rmann, J.~Hrubec, M.~Jeitler, W.~Kiesenhofer, M.~Krammer, D.~Liko, I.~Mikulec, M.~Pernicka, B.~Rahbaran, H.~Rohringer, R.~Sch\"{o}fbeck, J.~Strauss, A.~Taurok, F.~Teischinger, P.~Wagner, W.~Waltenberger, G.~Walzel, E.~Widl, C.-E.~Wulz
% ...
\vskip\cmsinstskip
% ...
1:~~Also at CERN, European Organization for Nuclear Research, Geneva, Switzerland\\
% ...

\end{sloppypar}
\end{document}